\documentclass[lettersize,journal]{IEEEtran}
\IEEEoverridecommandlockouts
\usepackage{cite}
\usepackage{amsmath,amssymb,amsfonts}
\usepackage{graphicx}
\usepackage{textcomp}
\usepackage{xcolor}
\usepackage{bm}
\usepackage{tabularx}
\usepackage{caption}
\usepackage{subcaption}
\usepackage{mathrsfs}
\usepackage{epstopdf}
\usepackage{textcomp}
\usepackage{hyperref}
\usepackage{float}  
\usepackage{diagbox} 
\usepackage{stfloats}
\usepackage{url}
\usepackage{verbatim}
\usepackage{algorithm}
\usepackage{algpseudocode}
\usepackage{titlesec}

\begin{document}

\title{CKMDiff: A Generative Diffusion Model for CKM Construction via Inverse Problems\\with Learned Priors
        \thanks{Part of this work will be presented at the IEEE VTC2025-Spring in Oslo, Norway, held from 17-20 June 2025.}
        \thanks{This research work is supported by Southeast University Network and Information Center.}
}

\author{Shen~Fu,~Yong~Zeng,~\IEEEmembership{Fellow,~IEEE},~Zijian~Wu,~Di~Wu,~Shi~Jin,~\IEEEmembership{Fellow,~IEEE},\\~Cheng-Xiang~Wang,~\IEEEmembership{Fellow,~IEEE},~and~Xiqi~Gao,~\IEEEmembership{Fellow,~IEEE}
\thanks{S. Fu, Y. Zeng, Z. Wu, D. Wu, S. Jin, C.-X. Wang, and X. Gao are with the National Mobile Communication Research Laboratory, School of Information Science and Engineering, Southeast University, Nanjing 211189, China. Y. Zeng, D. Wu, C.-X. Wang and X. Gao are also with Purple Mountain Laboratories, Nanjing 211111, China (e-mail: \{sfu, yong\_zeng, wuzijian, studywudi, jinshi, chxwang, xqgao\}@seu.edu.cn). (\textit{Corresponding author: Yong Zeng.})}
}
\maketitle
\begin{abstract}
    Channel knowledge map (CKM) is a promising technology to enable environment-aware wireless communications and sensing with greatly enhanced performance, by offering location-specific channel prior information for future wireless networks.
    One fundamental problem for CKM-enabled wireless systems lies in how to construct high-quality and complete CKM for all locations of interest, based on only limited and noisy on-site channel knowledge data. This problem resembles the long-standing ill-posed inverse problem, which tries to infer from a set of limited and noisy observations the cause factors that produced them.
    By utilizing the recent advances of solving inverse problems with learned priors using generative artificial intelligence (AI), we propose CKMDiff, a conditional diffusion model that can be applied to perform various tasks for CKM constructions such as denoising, inpainting, and super-resolution, without having to know the physical environment maps or transceiver locations.
    Furthermore, we propose an environment-aware data augmentation mechanism to enhance the model's ability to learn implicit relations between electromagnetic propagation patterns and spatial-geometric features.
    Extensive numerical results are provided based on the CKMImageNet and RadioMapSeer datasets, which demonstrate that the proposed CKMDiff achieves state-of-the-art performance, outperforming various benchmark methods. 

\end{abstract}


\section{Introduction}

The development of advanced wireless systems, particularly those employing extremely large-scale multiple-input multiple-output (XL-MIMO) \cite{luhaiquan,wcx1} and millimeter-wave (mmWave) \cite{wcx3} technologies, has heightened the need for efficient channel state information (CSI) acquisition in dynamic propagation environments.
However, obtaining accurate and timely CSI remains a significant challenge in these large-dimensional wireless systems, particularly in scenarios with high mobility or hardware limitations~\cite{csi3,csi2,wcx2}.
To address this challenge, recent studies~\cite{ckm,ckm1} have proposed the novel concept of channel knowledge map (CKM), which is a site-specific database, tagged with the locations of the transmitters and receivers, so as to provide location-specific channel prior information.
CKM enables a paradigm shift from the traditional environment-unaware wireless framework, which heavily relies on real-time channel estimation or environment sensing, to an environment-aware framework with learned a prior knowledge about the wireless environment.
From a communication perspective, CKM demonstrates particular efficacy in scenarios where real-time channel estimation is costly or even infeasible \cite{WDTRANS,WDWCL,DDY,QYL1,ZSQ,DZY1,comms1,comms2,comms3,comms4,comms5,comms6,comms8,comms10}, such as wireless communications involving non-cooperative nodes, high-dimensional antenna arrays, and passive devices like backscattering devices or reconfigurable intelligent surface (RIS).
For example, the authors in \cite{WDTRANS} proposed a CKM-assisted hybrid beamforming design scheme that leverages channel angle map (CAM) and beam index map (BIM) to enable environment-aware mmWave massive multiple-input multiple-output (MIMO) systems with significantly reduced training overhead.
In \cite{comms2}, CKM is utilized to optimize the placement of multiple unmanned aerial vehicles (UAVs) to maximize communication throughput.
Furthermore, the authors in \cite{comms4} combine CKM with historical CSI data for channel prediction to enhance beamforming performance in MIMO systems.
For integrated sensing and communication (ISAC) systems, CKM has also demonstrated great potential to enhance the sensing performance by offering channel prior information \cite{ZCY,isac1,isac2,xuzihan}.
For instance, recent work~\cite{xuzihan} proposed a novel clutter-suppression method by utilizing one novel CKM termed clutter angle map (CLAM), which provides location-specific angular information of the clutter so that spatial-domain clutter rejection can be used at the signal level.

However, before CKM can be widely adopted for wireless networks, its efficient construction methods need to be developed. One fundamental problem for CKM construction lies in how to utilize the limited and usually noisy on-site channel knowledge data to reconstruct a complete CKM for all locations of interest. In the extreme case, CKM may even need to be generated in the absence of any on-site channel knowledge data.
Note that on-site channel knowledge data is usually scarce due to the high cost of computational methods like ray tracing and the high expense of on-site measurements.
In fact, even ignoring the cost or expense issues, ray tracing and measurement methods can only acquire the channel knowledge data at discrete locations and discrete frequencies, which renders the aforementioned CKM construction problem essential.
Some preliminary research efforts have been devoted to developing various CKM construction methods, which can be broadly categorized into three types: channel model-based, interpolation-based, and artificial intelligence (AI)-based CKM construction, as discussed below.

A common approach is to leverage statistical channel model to construct CKMs.
For example, the authors in~\cite{xuxiaoli} introduced model-based spatial channel prediction for constructing channel gain map (CGM) from limited measurements, deriving theoretical relationships between data quantity (offline sampling density/online prediction points) and prediction accuracy to guide practical data collection for CKM construction.
A CKM construction method based on the expectation maximization (EM) algorithm was proposed in \cite{likun} by combining on-site channel gain data with expert knowledge derived from well-established statistical channel models. 
However, the quality of model-based CKM construction is biased by the pre-assumed mathematical models.
To achieve universal applicability, these models usually rely on simplistic assumptions and cannot accurately reflect the actual wireless channel in complex and diversified environments.

Another typical method for CKM construction is data interpolation or extrapolation, without assuming any pre-defined channel models. Typical methods include K-nearest neighbors (KNN)~\cite{knn}, Kriging~\cite{Kriging}, inverse distance weighting (IDW)~\cite{idw} and radial basis function (RBF)~\cite{RBF} interpolation.
However, interpolation-based methods exhibit inherent limitations since they rely solely on a limited number of surrounding observations to infer the unknown data points, without attempting to learn the intrinsic relations of the data.
Moreover, these approaches are inherently unable to account for the complex physical phenomena that drive channel variations, thereby restricting its efficacy in accurately modeling intricate environment.

With the recent advancement of AI, preliminary efforts have been made to develop AI-based CKM construction methods, which can be categorized into two types based on what conditions are provided for constructing CKM.
As demonstrated in \cite{I2I,radio_diff,construct1,construct2}, the first type focuses on mapping physical environment maps and/or transceiver location information to radio maps, which mainly provide path loss information.
For instance, the authors in~\cite{I2I} transformed the CKM construction problem into an image-to-image (I2I) inpainting task and proposed a deep-learning model, termed Laplacian Pyramid-based CGM Reconstructed Network (LPCGMN) to construct CKM from physical environment maps. 
The study~\cite{radio_diff} proposed a conditional diffusion model that generates radio maps in a sampling-free manner, utilizing physical maps and transmitter locations as conditional inputs.
The second category goes beyond relying solely on physical environment and/or transceiver location information, integrating partial observational data to build CKM.
For example, the UNet-based framework called RadioUNet in~\cite{radiounet} generates radio maps conditioned on sparse measurement points along with physical maps and transmitter positions.
For enhanced spatial resolution,~\cite{gan} implements a conditional generative adversarial networks (cGAN) to achieve radio map super-resolution using sparse measurements, physical maps and transceiver locations.
Although these methods demonstrate progress in CKM construction, they still require substantial prior knowledge of environmental geometry (e.g., complete geometric layouts, or transmitter locations), which may be unavailable in practical deployment scenarios.


To address the limitations of existing methods that rely on auxiliary environmental data and/or transceiver locations, we focus on reconstructing high-fidelity CKMs directly from limited and corrupted on-site observational data.
This is a challenging problem. Fortunately, it resembles the long-standing ill-posed inverse problem, where one seeks to infer causal factors from limited and usually noisy observations.
Generative AI, particularly diffusion models \cite{ddpm1}, provides a transformative framework for such problems by learning the prior information of the data to be reconstructed such as the underlying prior data distributions.
Motivated by this capability, we propose CKMDiff, a novel conditional diffusion model-based framework that effectively captures the intrinsic patterns of CKMs to enable simultaneous missing region recovery and noise perturbation suppression.
The key contributions are summarized as follows:
\begin{itemize}
    \item \textbf{CKM construction without requiring physical environment maps}: we propose a novel CKM construction method without relying on environmental information or transceiver location data, eliminating dependence on costly auxiliary measurements.
	\item \textbf{Multi-task, noise-robust CKM construction}: CKMDiff supports joint denoising, inpainting and super-resolution of CKMs by leveraging limited and noisy on-site channel knowledge data as conditional inputs to guide the diffusion process. Furthermore, CKMDiff is also applicable to an extreme scenario of CKM generation without any on-site channel knowledge data. 
    \item \textbf{Spatial-physical feature enhancement}: a novel feature emphasis mechanism is proposed to strengthen the model's capacity to learn spatially correlated channel patterns and implicit physical environment, improving fidelity in complex propagation environments.
	\item \textbf{Extensive evaluation and superior performance}: we conduct comprehensive evaluations on two datasets, RadioMapSeer~\cite{radiomapseer} and CKMImageNet~\cite{CKMImageNet}. Simulation results demonstrate that the proposed method outperforms existing baseline approaches in both super-resolution and inpainting tasks, showcasing its robustness and effectiveness across diverse scenarios.
\end{itemize}

The organization for the rest of this paper is as follows.
Section~\ref{sec:2} presents the system model for constructing CKMs from limited and noisy on-site observations.
Section~\ref{sec:3} reviews traditional methods that can be applied for CKM construction.
In Section~\ref{sec:4}, we introduce the theoretical foundations of denoising diffusion probabilistic models (DDPM) and latent diffusion models (LDM). We then present our conditional diffusion framework based on decoupled diffusion model (DDM) for CKM denoising, inpainting, and super-resolution without requiring physical environment maps, and describe our data augmentation strategy for enhancing CKM features.
Section~\ref{sec:5} elaborates on the characteristics of training datasets, simulation configurations and comprehensive simulation results validating the proposed approach.
Finally, Section~\ref{sec:6} draws the conclusion of this paper.

\emph{Notations:}
Scalars are denoted by italic letters.
Lower-case and upper-case letters in boldface represent vectors and matrices, respectively.
${\mathbb{R}^{M \times N}}$ denotes the space of $M \times N$ real-valued matrices.
$\mathbf{X}^T$, ${{\bf{X}}^{ - 1}} $, $\mathbf{X}^\dagger$ and ${\rm{Tr}}\left( {\bf{X}} \right)$ denote the transpose, inverse, pseudo-inverse and trace of matrix $\mathbf{X}$, respectively.
The $l_2$-norm of vector $\mathbf{x}$ is $\|\mathbf{x}\|$.
${\bf{I}}$ and ${\bf{0}}$ denote the identity matrix and zero matrix, respectively.
Sets are denoted by capital calligraphy letters.
${\cal N}(\mu,{\sigma ^2})$ denotes a Gaussian distribution with mean $\mu$ and variance $\sigma^2$, and $\sim $ stands for ``distributed as''.
Expectation is represented by ${\mathbb{E}}\left[  \cdot  \right]$. 


\section{System Model}
\label{sec:2}

\begin{figure}[t]
	\centerline{\includegraphics[width=9.5cm]{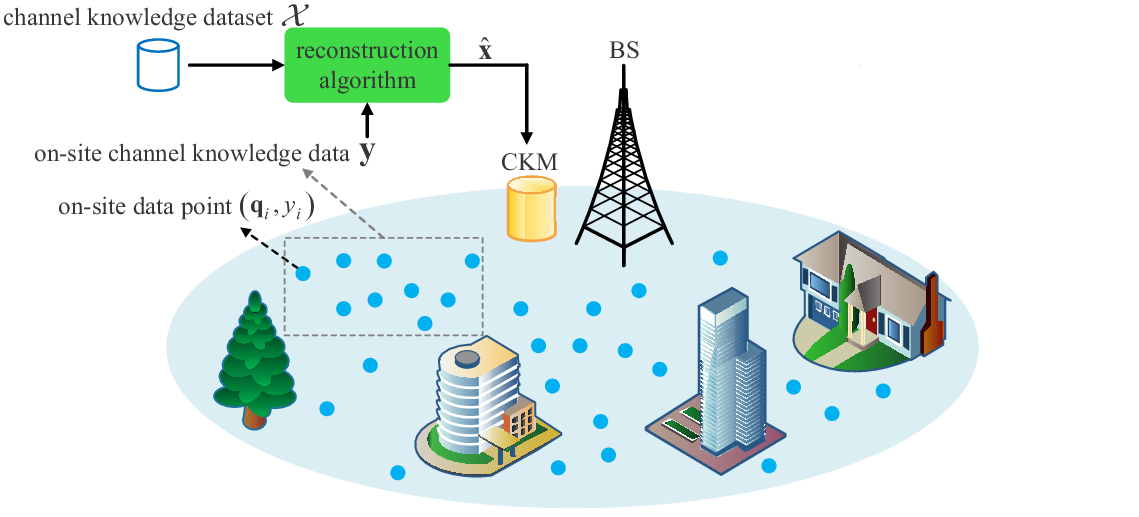}}
    \vspace{-4pt} 
	\caption{An Illustration for CKM construction based on limited and noisy on-site channel knowledge data.}
	\label{fig:scene}
\end{figure}


We consider the problem to construct a complete CKM for all locations in a given area, but only noisy on-site channel knowledge data is available for limited locations in this area.
This corresponds to various practical scenarios for CKM construction, where channel knowledge data is often incomplete and noisy due to high measurement or computational costs, as well as the inability to conduct on-site measurements in safety-restricted or inaccessible areas.
The area is partitioned into patches of equal size, each discretized into $l \times w$ grids.
Let ${\bf{x}} \in {\mathbb{R}^{{lw} \times 1}}$ denote the groundtruth of one type of channel knowledge at each of the $lw$ grids, which could be the channel gain, AoA, angle of departure (AoD), etc.  
The noisy and incomplete on-site channel knowledge observed can be represented as
\begin{equation}
    {\bf{y}} = {\bf{Ax}} + {\bf{n}},
    \label{equ:model}
\end{equation}
where ${\bf{y}} \in {\mathbb{R}^{{a} \times 1}}$ represents the observed channel knowledge, with $a\leq lw$,
and ${\bf{A}} \in {\mathbb{R}^{{a} \times {lw}}}$ denotes a known degradation matrix that varies with different tasks.
$\bf{n}$ is zero-mean Gaussian random noise, representing the noise introduced by various sources such as sensor imperfections, environmental interference, and measurement errors.

As shown in Fig.~\ref{fig:scene}, our objective is to reconstruct the complete CKM ${{\bf{ x}}}$ based on the observation ${\bf{y}}$ and ${{\bf{\hat x}}} \in {\mathbb{R}^{{lw} \times 1}}$ denotes the reconstructed CKM.
Let ${{y_i}}$ represent on-site channel knowledge data of the $i$th grid for ${\bf{y}}$ and ${{{\bf{q}}_i}} \in {\mathbb{R}^{{2} \times 1}}$ denote the corresponding 2-dimensional (2D) coordinates.
In addition, we collect a large off-line channel knowledge dataset ${\cal X}$.
Note that the mathematical model \eqref{equ:model} corresponds to a classic linear inverse problem that provides a unified model for several practical scenarios, which are illustrated in Fig.~\ref{fig:ckm_scenario} and discussed as follows:

\begin{figure}[t]
	\centerline{\includegraphics[width=9.0cm]{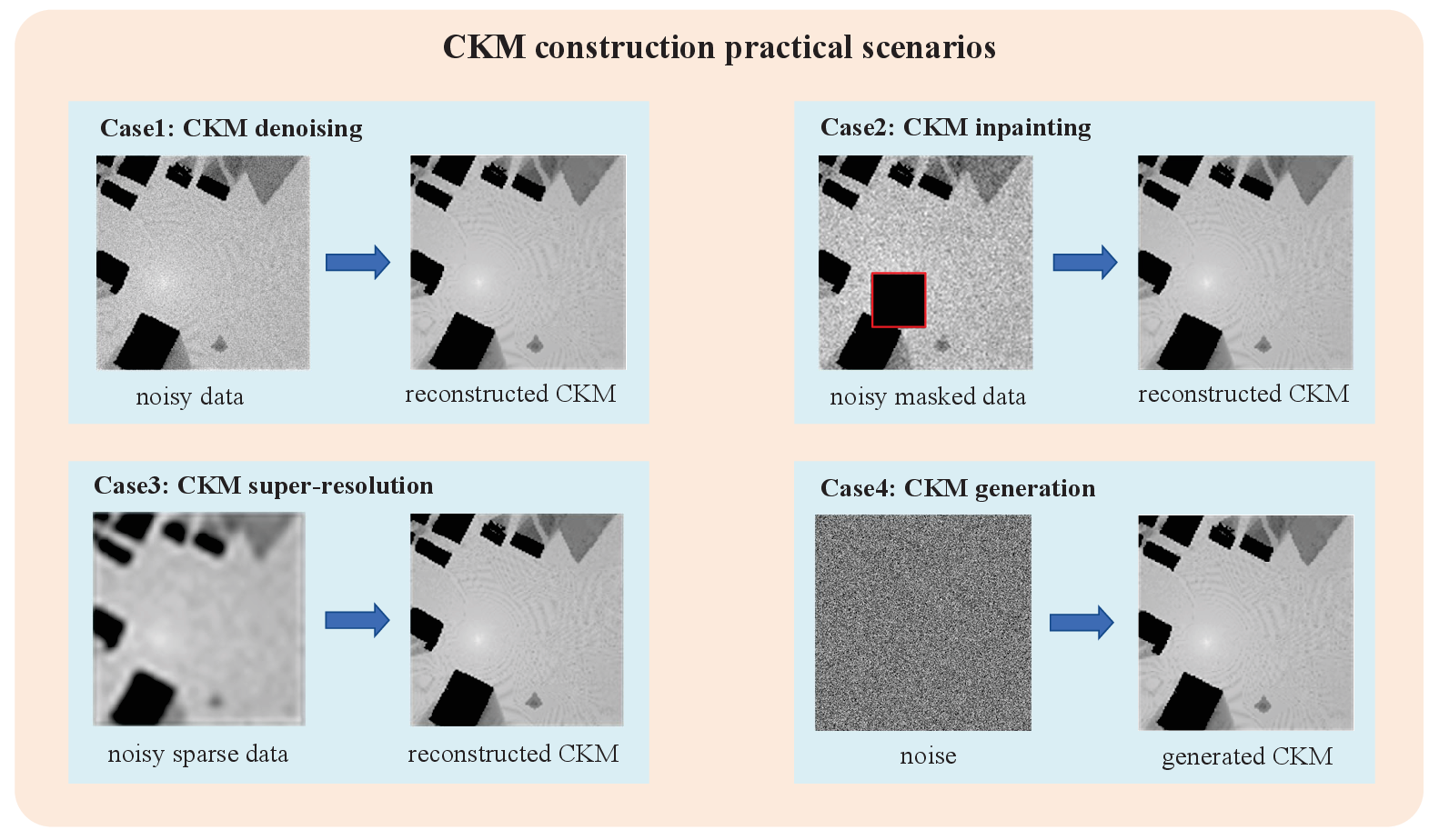}}
    \vspace{-4pt} 
	\caption{Various scenarios for CKM construction based on the unified model~\eqref{equ:model}.}
	\label{fig:ckm_scenario}
\end{figure}

(i) \textit{CKM denoising}: In this case, the observed channel knowledge data is complete, but it is corrupted by noise. 
As a result, we have $a=lw$ and ${\bf{A}} = {{\bf{I}}}$, so that ${\bf{y}} = {\bf{x}} + {\bf{n}}$.

(ii) \textit{CKM inpainting}:
If the channel knowledge data is missing in consecutive grids, then $a < lw$ and ${\bf{A}}$ is a full row rank matrix representing the mask with elements being zeros or ones.
The ${i}$th column of ${\bf{A}}$ is an all-zero vector if the channel knowledge at the corresponding grid $i$ is not observed.

(iii) \textit{CKM super-resolution}:
Sparse sampling strategies driven by practical constraints lead to a CKM super-resolution problem where high-resolution CKM ${{\bf{x}}}$ needs to be reconstructed from sparse observations ${\bf{y}}$.
In this case, ${\bf{y}} \in {\mathbb{R}^{{l'w'} \times 1}}$ denotes the low-resolution channel knowledge data, where $l' = {l \mathord{\left/ {\vphantom {l m}} \right. \kern-\nulldelimiterspace} m}$ and $w' = {w \mathord{\left/ {\vphantom {w m}} \right. \kern-\nulldelimiterspace} m}$ with $m$ being a downsampling factor.
${\bf{A}}$ is a downsampling matrix of size $l'w' \times lw$ which maps the high-resolution CKM ${\bf{x}}$ to a low-resolution observation ${\bf{y}}$.

(iv) \textit{CKM generation}: 
An extreme scenario in model \eqref{equ:model} is when $\bf{A}= {\bf{0}}$. 
In this case, the problem reduces to generating ${\bf{x}}$ solely from the Gaussian noise ${\bf{n}}$.
Thus, the CKM construction problem is transformed into a CKM generation problem.
This corresponds to the recently extensively studied generative AI framework \cite{ddpm1}, which is possible if the prior distributions of ${\bf{x}}$ is sufficiently learned. 

One important characteristic of the aforementioned problems with model \eqref{equ:model} is that information is incomplete or corrupted and perfect recovery of ${\bf{x}}$ is challenging. Specifically, when $a < lw$, the problem is ill-posed, i.e., there exist infinitely possible ${\bf{x}}$ that may explain the observation ${\bf{y}}$.
Fortunately, the recent advances of generative AI provides a novel mathematical framework to learn the prior distribution of the data ${\bf{x}}$. 
This makes it possible to regularize the ill-posed inverse problem by leveraging the learned prior distribution, thereby improving the accuracy and stability of the solution to the inverse problem \eqref{equ:model}.
Before presenting our proposed generative diffusion model-based CKM construction approach, we first review those traditional methods that may be applied to solve problem \eqref{equ:model} and expose their limitations.

\section{Conventional Methods}
\label{sec:3}
\subsection{Least Square-based CKM Construction}
For the linear inverse problem \eqref{equ:model}, one straightforward method to estimate ${\bf{x}}$ based on the observation ${\bf{y}}$ is to employ the least square (LS) method, which aims to minimize the squared error between the observed data  ${\bf{y}}$ and predicted data ${\bf{Ax}}$. 
The cost function to be minimized is given by: 
\begin{equation}
    J\left( {\bf{x}} \right) = {\left\| {{\bf{y}} - {\bf{Ax}}} \right\|^2}.
    \label{ls}
\end{equation}
Accordingly, the LS solution is
\begin{equation}
    {\bf{\hat x}} = {{\bf{A}}^\dag }{\bf{y}},
\end{equation}
where ${{\bf{A}}^\dag } = {{\bf{A}}^T}{\left( {{\bf{A}}{{\bf{A}}^T}} \right)^{ - 1}}$ is the pseudo-inverse of ${\bf{A}}$.

The LS approach faces significant challenges in CKM construction.
First, LS lacks regularization, making it highly sensitive to noise. In scenarios where $a < lw$, noise is amplified, leading to unphysical artifacts such as oscillatory patterns in the reconstructed CKM.
Second, LS tries to minimize the prediction error without using any prior information about the data $\bf{x}$, such as its mean or correlations, let alone the actual prior distribution of data. 
This renders it usually achieve poor performance, especially when the observation is very sparse $a \ll lw$ and/or noise is severe.

\subsection{Interpolation-based CKM Construction}
Interpolation methods are widely used to estimate missing data based on observed sparse data. 
A unified formula for linear interpolation-based CKM construction is:
\begin{equation}
    {\hat x_i} = \sum\limits_{j \in {\cal S}_i} {{w_j}{y_j}},
    \label{interpolation}
\end{equation}
where ${\hat x_i}$ is the predicted channel knowledge at grid $i$, ${y_j}$ denotes the observed channel knowledge at grid $j$,
${{\cal S}_i}$ is the set of observed channel knowledge used for interpolating target channel knowledge at grid $i$, and ${w_j}$ represents the weight.

Different interpolation methods vary primarily in how to determine the set ${{\cal S}_i}$, the weights ${w_j}$, and the underlying assumptions about spatial relationships.
KNN~\cite{knn} predicts a missing pixel ${x_i}$ in ${\bf{x}}$ by averaging the values of its $ K$ nearest neighbors in $\bf{y}$, where ${w_j} = {1 \mathord{\left/ {\vphantom {1 K}} \right. \kern-\nulldelimiterspace} K}$ and ${{\cal S}_i}$ is the set of $K$ nearest channel knowledge data for ${x_i}$.
Bilinear~\cite{bilinear} estimates a missing pixel ${x_i}$ in ${\bf{x}}$ by linearly combining its four nearest neighbors in $ \bf{y}$, and thus ${{\cal S}_i}$ is the set of four nearest observed channel knowledge data for ${x_i}$ and ${{\omega _j}}$ are determined by the relative distances between the target pixel ${{x_i}}$ and known pixels ${{y_j}}$.
IDW~\cite{idw} assigns weights inversely proportional to the distance between the target channel knowledge data and observed channel knowledge data. 
Kriging~\cite{Kriging,kriging1} calculates weights based on a variogram model, which describes the spatial correlation between observed data, and the weights are optimized to minimize the variance of estimation error.




Since CKM often exhibits complex spatial patterns and nonlinear dependencies, which cannot be reflected by interpolation methods due to their assumption of simple spatial continuity.
Additionally, interpolation struggles to capture rapid signal variations and is highly sensitive to noise, which is common in real-world channel knowledge data. 

\subsection{Spatial Correlation Model-based CKM Construction}

Spatial correlation model-based CKM construction usually utilizes a certain statistical channel model to capture the channel knowledge correlations at different locations.
Based on observed channel knowledge data, the model parameters ${\bm{\rho }}$ are estimated via curve fitting techniques.
These fitted parameters are then utilized to predict channel knowledge at unknown locations, which can be expressed as:
\begin{equation}
    {{\hat x}_i} = f\left( {{{\bf{y}}_i},{\bm{\rho }}} \right),
    \label{equ:model_CGM}
\end{equation}
where ${{\hat x}_i}$ is the estimated channel data in position $i$, $f\left(  \cdot  \right) $ is a function that incorporates the statistical channel model,
and ${{\bf{y}}_i}$ denotes the channel knowledge data at the $k$ nearest observed points for unknown position $i$. 

For instance, the authors in \cite{xuxiaoli} used the classic path loss model to complete CGM. 
The channel gain consists of three major components, i.e., the path loss, shadowing and multipath fading, and it is modelled by
\begin{equation}
    {\Upsilon _{dB}}\left( {\bf{q}} \right) = {K_{dB}} - 10{n_{PL}}{\log _{10}}\left( {\left\| {\bf{q}} \right\|} \right) + \nu \left( {\bf{q}} \right) + \omega \left( {\bf{q}} \right),
    \label{equ:model_gain}
\end{equation}
where ${\Upsilon _{dB}}\left( {\bf{q}} \right)$ denotes the channel gain in dB between the BS located at the origin and a receiver at location ${\bf{q}}$.
${K_{dB}}$ represents the path loss intercept, and $n_{PL}$ is the path loss exponent.
The shadowing effect $\nu \left( {\bf{q}} \right)$ is modelled as a zero-mean Gaussian random variable with variance $o $ and spatial correlation function
\begin{equation}
    \varepsilon \left( h \right) = o\exp \left( { - \frac{h}{d}} \right),
    \label{equ:spatial}
\end{equation}
where $h$ denotes the distance between the pair of data points and $d $ is known as the correlation distance of shadowing.
$\omega \left( {\bf{q}} \right)$ denotes the effect of multipath and is modelled a zero-mean Gaussian random variable, with variance ${\sigma ^2}$.
The authors first adopt the LS estimation to estimate the path loss intercept ${{\hat K}_{dB}}$ and path loss exponent ${{\hat n}_{PL}}$. 
The residual signal can be obtained by subtracting the estimated path loss from the measured channel gain, which captures both shadowing and fading effects.
Analyzing the spatial correlation of these residuals enables the estimation of the shadowing variance ${\hat o}$, correlation distance ${\hat d}$, and the multipath fading variance ${{\hat \sigma }^2}$.




Spatial correlation model-based CKM construction typically relies on predefined models that incorporate simplifying assumptions, such as uniform propagation environments or idealized geometric configurations.
However, real-world electromagnetic propagation environments are inherently complex and non-uniform.
It is challenging for any single model to fully and accurately describe the true propagation characteristics, leading to model-induced biases in the recovery performance.

\subsection{Prior Distribution-based CKM Construction}
Another typical approach for CKM construction is to leverage the full prior distribution $p\left( {\bf{x}} \right)$ of the data $\bf{x}$ to be reconstructed.
In particular, maximum a posteriori (MAP) estimation reconstructs the complete CKM $\bf{x}$ from observations $\bf{y}$ by solving:
\begin{equation}
    {\bf{\hat x}} = \arg \mathop {\max }\limits_{\bf{x}} p\left( {{\bf{x}}\left| {\bf{y}} \right.} \right) = \arg \mathop {\max }\limits_{\bf{x}} p\left( {{\bf{y}}\left| {\bf{x}} \right.} \right)p\left( {\bf{x}} \right),
    \label{equ:map}
\end{equation}
where $p\left( {{\bf{y}}\left| {\bf{x}} \right.} \right)$ is the likelihood function and $p\left( {\bf{x}} \right)$ is the prior distribution of channel data.
MAP seeks the most probable reconstruction given the data and prior.

On the other hand, minimum mean square error (MMSE) estimation computes the mean of the posterior distribution:
\begin{equation}
    {\bf{\hat x}} = \mathbb{E}\left[ {{\bf{x}}\left| {\bf{y}} \right.} \right] =\int {{\bf{x}}p\left( {{\bf{x}}\left| {\bf{y}} \right.} \right)d{\bf{x}}} ,
    \label{equ:MMSE}
\end{equation}
which minimizes the expected squared error $\mathbb{E}{\left\| {{\bf{x}} - {\bf{\hat x}}} \right\|^2}$.
While MMSE is optimal in a mean-squared sense, it inherently averages over all plausible solutions weighted by their posterior probabilities, which is known to suffer from the \textit{regression to the mean} issue.


Unlike generative AI that produces diverse plausible CKM realizations, MAP and MMSE only provide a single-point estimate, failing to capture the intrinsic uncertainty in wireless environments.
Besides, the performance of MAP and MMSE critically depends on accurate knowledge of $p\left( {\bf{x}} \right)$ or $p\left( {{\bf{x}}\left| {\bf{y}} \right.} \right)$, yet closed-form modeling of these distributions for realistic CKM deployments is mathematically intractable. 

\section{Proposed CKM Construction via Diffusion Models}
\label{sec:4}

The CKM construction problem in \eqref{equ:model} is analogous to image restoration problems.
Therefore, we propose leveraging generative AI techniques from image processing to learn the prior distribution of channel knowledge data. To enable AI-driven CKM construction, we have collected a large channel knowledge dataset ${\cal X} = \left\{ {{{\bf{x}}^1},{{\bf{x}}^2}, \cdots ,{{\bf{x}}^P}} \right\}$, where $P$ denotes the size of the dataset.
In particular, diffusion models~\cite{ddpm1,ddpm} have recently gained popularity due to their training stability and high-quality generation capabilities, outperforming generative adversarial networks (GANs)~\cite{GAN2} and other alternatives.
Building on this, we leverage a conditional diffusion model-based approach for CKM construction, aiming to capture the intrinsic relationships and latent structures within channel knowledge data.

\subsection{Data Pre-processing}

In order to treat channel knowledge data as images, it is often more convenient for the inpainting task to make the observation ${\bf{y}}$ to have equal size as ${\bf{x}}$, i.e., ${a = lw}$.
To this end, we fill the unobserved positions in the channel knowledge data with zeros.
Consequently, the processed observed data can be expressed as follows:
\begin{equation}
    {\bf{\tilde y}} = {{{\bf{A}}^T}\bf{y}},
\end{equation}
where ${\bf{\tilde y}} \in {\mathbb{R}^{{lw} \times 1}}$ denotes the post-processed observation and can be treated as an image of size $l \times w$.
For the super-resolution task, the observation ${\bf{y}}$ is viewed as a low-resolution image with size $l' \times w'$. 

\subsection{DDPM}


The core principle of DDPM proposed in \cite{ddpm1}, is to learn a generative distribution ${p_{\bf{\theta }}}\left( {{\bf{x}}_0} \right)$ by minimizing a variational bound on the negative log-likelihood, such that ${p_{\bf{\theta }}}\left( {{\bf{x}}_0} \right)$ approximates the true data distribution $q\left( {{\bf{x}}_0} \right)$,
where ${{\bf{x}}_0}$ denotes the clean data at forward diffusion step $0$ (i.e., the original real CKM).
DDPM utilizes a forward process to gradually corrupt the data by adding Gaussian noise over a sequence of time steps $t \in \left[ {1,T} \right]$, and a reverse process to reconstruct the original data by denoising step-by-step.
Starting from a clean image ${{\bf{x}}_0}$, the forward process is defined as:
\begin{equation}
	q\left( {{{\bf{x}}_t}|{{\bf{x}}_{t - 1}}} \right) = {\cal N}\left( {{{\bf{x}}_t};\sqrt {1 - {\beta _t}} {{\bf{x}}_{t - 1}},{\beta _t}{\bf{I}}} \right),
	\label{forward process}
\end{equation}
where ${{\bf{x}}_t}$ represents the noised version of the original data ${{\bf{x}}_0}$ after diffusion steps $t$ and ${\beta _t}$ is a variance schedule that determines how much noise is added at each step $t$.
The process is continued till time step $T$, when ${{\bf{x}}_0}$ is completely degraded to noise.
It is known that the forward process admits sampling ${{\bf{x}}_t}$ at an arbitrary timestep $t$ in a closed form:
\begin{equation}
	q\left( {{{\bf{x}}_t}|{{\bf{x}}_0}} \right) = {\cal N}\left( {{{\bf{x}}_t};\sqrt {{{\bar \alpha }_t}} {{\bf{x}}_0},\left( {1 - {{\bar \alpha }_t}} \right){\bf{I}}} \right),
	\label{forward x0}
\end{equation}
where ${{\bar \alpha }_t} = \prod\nolimits_{s = 1}^t {{\alpha _s}} $ and ${{\alpha _t} = 1 - {\beta _t}}$. 
Accordingly, we can sample ${{\bf{x}}_t}$ by ${{\bf{x}}_t} = \sqrt {{{\bar \alpha }_t}} {{\bf{x}}_0} + \sqrt {1 - {{\bar \alpha }_t}} {\bm{\epsilon}}$ with $\bm{\epsilon}  \sim {\cal N}\left( {0,{\bf{I}}} \right)$.

The reverse process aims to recover the original data by eliminating  the added noise gradually.
By defining a Markov chain, it starts with the highly noisy data ${{\bf{x}}_T}$ and moves in reverse to ${{\bf{x}}_0}$. The specific process from ${{\bf{x}}_t}$ to ${{\bf{x}}_{t-1}}$ can be expressed by
\begin{equation}
	{p_{\bm{\theta}}}\left( {{{\bf{x}}_{t - 1}}|{{\bf{x}}_t}} \right) = {\cal N}\left( {{{\bf{x}}_{t - 1}};{{\bm{\mu}}_{\bm{\theta}} }\left( {{{\bf{x}}_t},t} \right),{{\bm{\Sigma }}_{\bm{\theta}} }\left( {{{\bf{x}}_t},t} \right)} \right).
	\label{reverse process}
\end{equation}
Here, ${{\bm{\mu} _{\bm{\theta}} }}$ and ${{\bm{\Sigma} _{\bm{\theta}} }}$ are learned functions parameterized by ${\bm{\theta}}$,
which correspond to the mean and the covariance matrix of the assumed Gaussian distribution.
According to \cite{ddpm1}, it is proved that that the reverse process can learn ${{\bm{\mu}}_{\bm{\theta}} }\left( {{{\bf{x}}_t},t} \right)$.
When setting ${{\bm{\Sigma }}_{\bm{\theta}} }\left( {{{\bf{x}}_t},t} \right) = \sigma _t^2{\bf{I}}= \frac{{1 - {{\bar \alpha }_{t - 1}}}}{{1 - {{\bar \alpha }_t}}}{\beta _t}{\bf{I}}$,
${{\bm{\mu}}_{\bm{\theta}} }\left( {{{\bf{x}}_t},t} \right)$ is given by
\begin{equation}
    {{\bm{\mu}}_{\bm{\theta}} }\left( {{{\bf{x}}_t},t} \right) = \frac{1}{{\sqrt {{\alpha _t}} }}\left( {{{\bf{x}}_t} - \frac{{{\beta _t}}}{{\sqrt {1 - {{\bar \alpha }_t}} }}{\bm{\epsilon} _\theta }\left( {{{\bf{x}}_t},t} \right)} \right),
    \label{equa:mean}
\end{equation}
where ${\bm{\epsilon} _\theta }$ is a function approximator intended to predict ${\bm{\epsilon}}$ from ${{\bf{x}}_t}$.
Therefore, we main train a neural network to minimize the following loss function of DDPM:
\begin{equation}
    {{\cal L}_{DDPM}} =  {\mathbb{E}_{{\bm{\epsilon}\sim {\cal N}\left( {0,{\bf{I}}} \right)},t,{{\bf{x}}_0}}}\left[ {{{\left\| {{\bm{\epsilon }} - {{\bm{\epsilon }}_{\bm{\theta}} }\left( {{{\bf{x}}_t},t} \right)} \right\|}^2}} \right].
	\label{loss function}
\end{equation}

After training the denoiser network ${{\bm{\epsilon }}_{\bm{\theta}} }$, sampling generates data by progressively denoising from pure Gaussian noise:
\begin{equation}
	{{\bf{x}}_{t - 1}} = \frac{1}{{\sqrt {{\alpha _t}} }}\left( {{{\bf{x}}_t} - \frac{{1 - {\alpha _t}}}{{\sqrt {1 - {{\bar \alpha }_t}} }}{{\bm{\epsilon }}_\theta }\left( {{{\bf{x}}_t},t} \right)} \right) + {\sigma _t}{\bf{p}},
	\label{sample}
\end{equation}
where $\bf{p}  \sim {\cal N}\left( {0,{\bf{I}}} \right)$.

\subsection{LDM}
The aforementioned DDPM operates directly in the high-dimensional pixel space, making them computationally expensive, especially when dealing with large or high-resolution datasets.
To address these challenges, LDM proposed in \cite{ldm} can be applied, where the diffusion process is applied in a lower-dimensional latent space. 
LDM first encodes the input data ${{\bf{x}}_0}$ into a compact latent representation ${{\bf{z}}_0}$ using a pre-trained variational autoencoder (VAE) \cite{vae}, where ${{\bm{{\cal E}}}}$ and ${\bm{{\cal D}}}$ denote the encoder and decoder of VAE, respectively.
The diffusion process is then applied to this latent space, which is typically much smaller than the original pixel space.
This allows the model to operate more efficiently, reducing both the computational cost and memory requirements.


The cost function for a trained neural network to minimize is represented as
\begin{equation}
    {{\cal L}_{LDM}} =  {\mathbb{E}_{t,{{\bm{{\cal E}}}\left( {{\bf{x}}_0} \right)},{\bm{\epsilon }\sim {\cal N}\left( {0,{\bf{I}}} \right)}}}\left[ {{{\left\| {{\bm{\epsilon }} - {{\bm{\epsilon }}_{\bm{\theta}} }\left( {{{\bf{z}}_t},t} \right)} \right\|}^2}} \right],
	\label{lossldm}
\end{equation}
where ${{\bf{z}}_t}$ represents the noised version of the original latent data ${{\bf{z}}_0}$ after diffusion steps $t$ and can be efficiently obtained from encoder ${\bm{{\cal E}}}$ during training since the forward process is fixed, and samples from $p\left( {\bf{z}} \right)$ can be decoded to image pixel space with a single pass through pre-trained decoder ${\bm{{\cal D}}}$.

\subsection{Conditional DDM for CKM Construction}
\begin{figure*}[t]
	\centerline{\includegraphics[width=14.5cm]{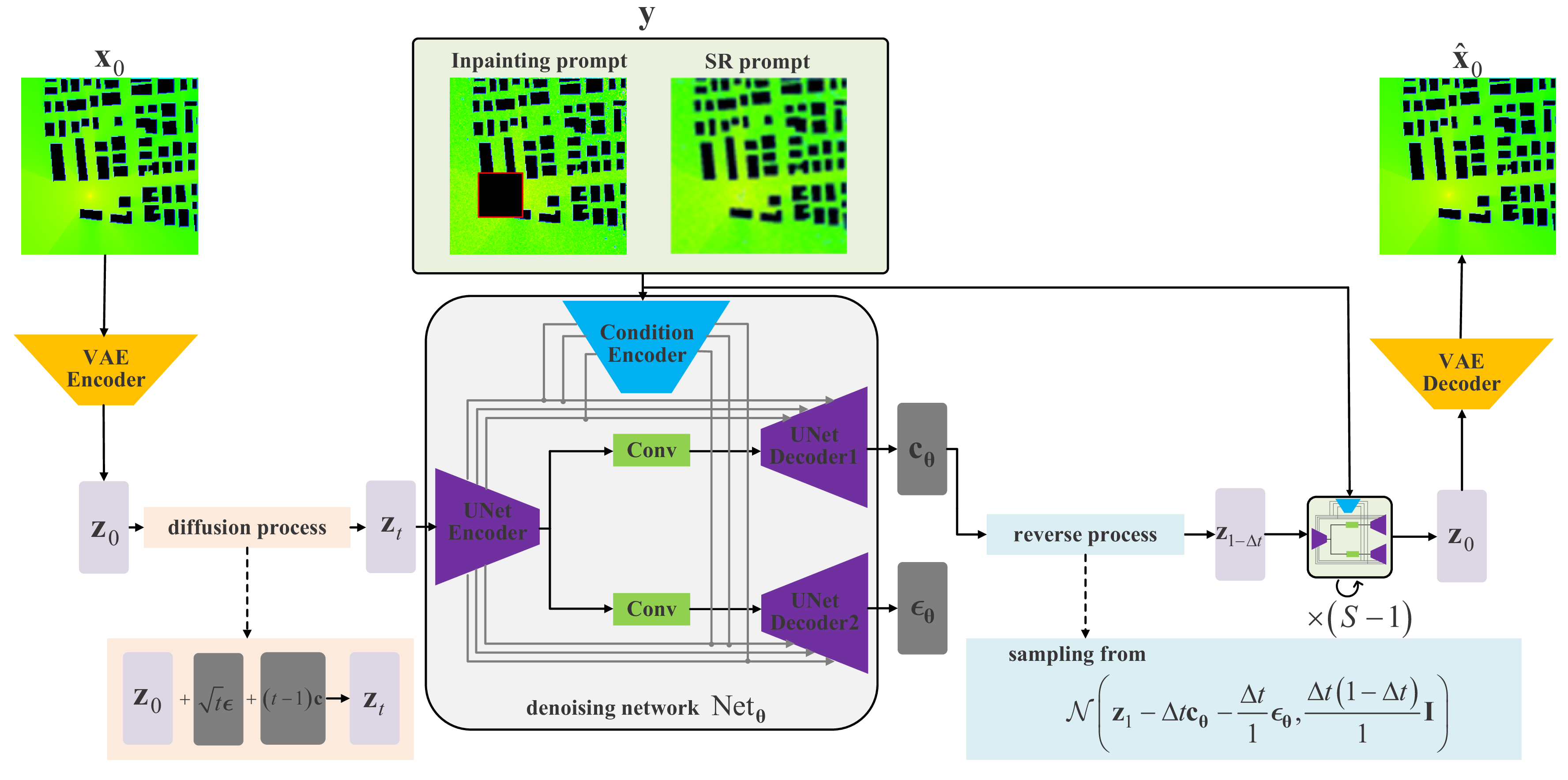}}
    \vspace{-4pt} 
	\caption{Conditional decoupled diffusion model-based CKM construction architecture.}
	\label{fig:ddm}
\end{figure*}

In order to further improve the quality and speed of CKM reconstruction, we propose to use the conditional DDM \cite{ddm} to solve the inverse problem \eqref{equ:model}. 
Different from DDPM, DDM converts the normal image-to-noise forward diffusion process into two stages:(i) attenuate the image component by an image-to-zero mapping, (ii) increase the noise component by a zero-to-noise mapping.
By utilizing decoupled diffusion strategy, the forward diffusion process is represented by \cite{ddm}
\begin{equation}
	{{\bf{x}}_t} = {{\bf{x}}_0} + \int_0^t {{{\bf{f}}_\tau }d\tau}  + \int_0^t {d{{\bf{w}}_\tau}},
	\label{forward1 ddm}
\end{equation}
where ${{\bf{x}}_0} + \int_0^t {{{\bf{f}}_\tau}d\tau}$ denotes the image attenuation process and $\int_0^t {d{{\bf{w}}_\tau}}$ represents the noise enhancement process. 
${{\bf{w}}_t}$ is the standard Wiener process, and ${{{\bf{f}}_t}}$ is a differentiable function of ${t}$ that allows arbitrary sampling steps, thus accelerating the sampling process. 
In this paper, we set ${{{\bf{f}}_t}}$ as a simple but effective function ${{{\bf{f}}_t} = \bf{c}}$.
Note that $t$ takes the range $\left[ {0,1} \right]$, different from DDPM. 
To achieve the forward diffusion process from ${{\bf{x}}_0}$ to noise, we need to ensure that the transformation effectively converts the initial state into a noise-dominated state.
Thus, ${{\bf{x}}_0} $ follows the distribution $q\left( {{{\bf{x}}_0}} \right)$ and ${{\bf{x}}_0} + \int_0^1 {{{\bf{f}}_\tau}d\tau}  = {\bf{0}}$, ensuring that ${{\bf{x}}_1}$ is distributed as ${\cal N}\left( {{\bf{0}},{\bf{I}}} \right)$. 
In other words, ${\bf{c}} =  - {{\bf{x}}_0}$.

As a result, the forward process can be rewritten as:
\begin{equation}
    q\left( {{{\bf{x}}_t}|{{\bf{x}}_0}} \right) = {\cal N}\left( {{{\bf{x}}_t};\left( {1 - t} \right){{\bf{x}}_0},t{\bf{I}}} \right).
	\label{forward ddm}
\end{equation}
Therefore, we can sample ${{\bf{x}}_t}$ by ${{\bf{x}}_t} = \left( {1 - t} \right){{\bf{x}}_0} + \sqrt t {\bm{\epsilon }}$, where ${\bm{\epsilon }} \sim {\cal N}\left( {{\bf{0}},{\bf{I}}} \right)$. 

Unlike DDPM which uses discrete time Markov chain, the reverse process in DDM employs continuous-time Markov chain with the smallest time step ${\Delta t \to {0^ + }}$ and we use conditional distribution ${q\left( {{{\bf{x}}_{t - \Delta t}}|{{\bf{x}}_t},{{\bf{x}}_0}} \right)}$ to approximate ${q\left( {{{\bf{x}}_{t - \Delta t}}|{{\bf{x}}_t}} \right)}$:
\begin{equation}
\small
	q\left( {{{\bf{x}}_{t - \Delta t}}|{{\bf{x}}_t},{{\bf{x}}_0}} \right) = {\cal N}\left( {{{\bf{x}}_t} - \Delta t{\bf{c}} - \frac{{\Delta t}}{{\sqrt t }}{\bm{\epsilon }},\frac{{\Delta t\left( {t - \Delta t} \right)}}{t}{\bf{I}}} \right).
    \label{reverse ddm}
\end{equation}


From the reverse process in \eqref{reverse ddm}, signal attenuation term $- \Delta t{\bf{c}}$ and noise term ${{\bm{\epsilon }}}$ are unknown. 
Hence, we use ${p_{\bm{\theta }}}\left( {{{\bf{x}}_{t - \Delta t}}|{{\bf{x}}_t}} \right)$ to approximate $q\left( {{{\bf{x}}_{t - \Delta t}}|{{\bf{x}}_t},{{\bf{x}}_0}} \right)$ and simultaneously predict ${\bf{c}}$ and ${{\bm{\epsilon }}}$ by utilizing a modified UNet architecture ${\rm{Net}}_{\bm{\theta}}$, where two stacked convolutional layers are added to create two UNet decoder branches.
Besides, like LDM, the encoder of a pretrained VAE ${\bm{{\cal E}}}$ is employed to map the pixel space ${{\bf{x}}_0}$ to the latent space ${{\bf{z}}_0}$.
The decoder $\bm{{\cal D}}$ is then used to map the latent space ${{\bf{\hat z}}_0}$ back to the pixel space ${{\bf{\hat x}}_0}$, reconstructing the input data while preserving the essential features learned during the encoding process.

To address CKM construction problem with partially-observed data, a Swin-B encoder $e$ \cite{swin} is used to extract multi-level features of the conditioned input, i.e., the observed channel knowledge data ${\bf{y}}$, and concatenate these features with the image features at the UNet decoder's input levels.
Therefore, $ {p_{\bm{\theta }}}\left( {{{\bf{z}}_{t - \Delta t}}|{{\bf{z}}_t},e\left( {\bf{y}} \right)} \right)$ can be expressed as

\begin{equation}
    \begin{aligned}
    p_{\bm{\theta}}\left( {\bf{z}}_{t - \Delta t} \mid {\bf{z}}_t, e({\bf{y}}) \right) = & \mathcal{N}\left( {\bf{z}}_{t - \Delta t}; {\bf{z}}_t - \Delta t {\bf{c}}_{\bm{\theta}} \left( {\bf{z}}_t, t \mid e({\bf{y}}) \right) \right. \\
    & \left. - \frac{\Delta t}{\sqrt{t}} \bm{\epsilon}_{\bm{\theta}} \left( {\bf{z}}_t, t \mid e({\bf{y}}) \right), \frac{\Delta t (t - \Delta t)}{t} \mathbf{I} \right).
    \end{aligned}
\end{equation}

Accordingly, the training cost function is expressed as


\begin{equation}
\resizebox{0.98\linewidth}{!}{$\displaystyle 
\mathop {\min }\limits_{\bm{\theta}}  
{\mathbb{E}_{t,{{\bm{{\cal E}}}\left( {{\bf{x}}_0} \right)},{\bm{\epsilon }},{\bf{y}}}}
\left[ {{{\left\| {{\bf{c}} - {{\bf{c}}_{\bm{\theta}} }\left( {{{\bf{z}}_t},t|e\left( {\bf{y}} \right)} \right)} \right\|}^2} + {{\left\| {{\bm{\epsilon }} - {{\bm{\epsilon }}_{\bm{\theta}} }\left( {{{\bf{z}}_t},t|e\left( {\bf{y}} \right)} \right)} \right\|}^2}} \right].
$}
\label{ddm loss}
\end{equation}

To sum up, the specfic training and sampling procedures for the proposed CKM construction algorithm are summarized in Alg. \ref{train_algorithm} and Alg. \ref{sample_algorithm}, respectively.
The architecture of the neural network is demonstrated in Fig. \ref{fig:ddm}. 

\begin{algorithm}[t]
    \caption{The training algorithm for CKM construction}
     \begin{algorithmic}[1]
        \State \textbf{Input}: channel knowledge dataset following the unknown distribution ${{\bf{x}}_0} \sim q({{\bf{x}}_0})$, pre-trained VAE encoder ${\bm{{\cal E}}}$, condition encoder $e$
        \State \textbf{Output}: trained network ${\rm{Net}}_{\bm{\theta}}$
        \State \textbf{Initialize}: $i = 0$, number of iterations: $N$, network parameters: ${\bm{\theta }}$
        \While{$i < N$}
            \State sample ${{\bf{x}}_0} \sim q({{\bf{x}}_0})$ and generate the corrupted observation ${\bf{y}} = {\bf{A}}{{\bf{x}}_0} + {\bf{n}}$, where ${\bf{A}}$ is a random degradation matrix. sample $t \sim \text{Uniform}(0, 1)$, $\bm{\epsilon} \sim \mathcal{N}({\bf{0}}, {\bf{I}})$
            \State ${{\bf{z}}_0} = {\bm{{\cal E}}}\left( {{\bf{x}}_0} \right)$,  ${\bf{c}}= - {{\bf{z}}_0}$
            \State ${\bf{z}}_t = {\bf{z}}_0 + t {\bf{c}} + \sqrt{t} \bm{\epsilon}$
            \State ${\bf{c}}_{\bm{\theta}}, \bm{\epsilon}_{\bm{\theta}} = \text{Net}_{\bm{\theta}}({\bf{z}}_t, t, e({\bf{y}}))$
            \State \textbf{Take gradient descent step on}
            
            ${\nabla_{\bm{\theta}}} \left( \left\| {\bf{c}} - {\bf{c}}_{\bm{\theta}}({\bf{z}}_t, t | e({\bf{y}})) \right\|^2 + \left\| \bm{\epsilon} - \bm{\epsilon}_{\bm{\theta}}({\bf{z}}_t, t | e({\bf{y}})) \right\|^2 \right)$
            \State $i = i + 1$
        \EndWhile
        \State \textbf{return} $\bm{\theta}$
    \end{algorithmic}
    \label{train_algorithm}
    \end{algorithm}
    
    \begin{algorithm}[t]
    \caption{The sampling algorithm for CKM construction}
    \begin{algorithmic}[1]
        \State \textbf{Input}: degraded channel knowledge observation ${\bf{y}}$, trained network ${\rm{Net}}_{\bm{\theta}}$, pre-trained VAE decoder ${\bm{{\cal D}}}$, condition encoder $e$  
        \State \textbf{Output}: constructed CKM ${{\bf{\hat x}}_0}$
        \State \textbf{Initialize}: ${\bf{z}}_1 \sim \mathcal{N}({\bf{0}}, {\bf{I}})$, number of sampling steps: $S$, sampling interval: $\Delta t = 1/S$
        \While{$t > 0$}
            \State ${\bm{\tilde{\epsilon}}} \sim \mathcal{N}({\bf{0}}, {\bf{I}})$
            \State ${\bf{c}}_{\bm{\theta}}, {\bm{\epsilon}}_{\bm{\theta}} = \text{Net}_{\bm{\theta}}({\bf{z}}_t, t, e({\bf{y}}))$
            \State ${\bf{z}}_t = {\bf{z}}_t - \Delta t {\bf{c}}_{\bm{\theta}} - \frac{\Delta t}{\sqrt{t}} {\bm{\epsilon}}_{\bm{\theta}} + \sqrt{\frac{\Delta t(t - \Delta t)}{t}} \bm{\tilde{\epsilon}}$
            \State $t = t - \Delta t$
        \EndWhile
        \State ${{\bf{\hat x}}_0} = {\bm{{\cal D}}}\left( {{{\bf{z}}_0}} \right)$
        \State \textbf{return} ${{\bf{\hat x}}_0}$
    \end{algorithmic}
    \label{sample_algorithm}
\end{algorithm}

\subsection{Data Feature Emphasis}
\begin{figure}[t]
	\centerline{\includegraphics[width=8.5cm]{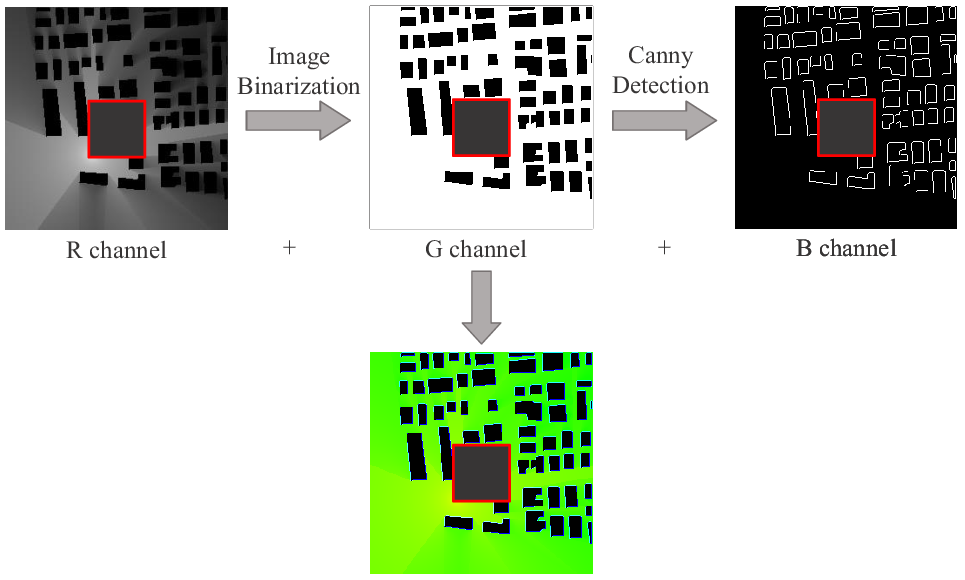}}
    \vspace{-4pt} 
	\caption{CKM data feature extraction.}
	\label{fig:feature}
\end{figure}

Our preliminary work \cite{me}, which utilized CKM gray images directly as input to the diffusion model for CKM construction, aimed to leverage the inherent characteristics of CKM to model and predict wireless propagation properties in a given environment.
In this work, we jointly learn channel knowledge data, the distribution of buildings in the environment, and the boundaries of these structures.
Unlike previous CKM construction studies \cite{I2I,radio_diff,radiounet,gan,construct2,construct1}, which require complete physical environment maps, our method does not rely on prior knowledge of the full physical environment.
Instead, we focus on extracting features from incomplete channel data, as illustrated in Fig.~\ref{fig:feature}. These features are transformed into an RGB-like representation, which serves as the input to our model.
This approach enables CKM construction by implicitly capturing the spatial and structural relationships between channel properties and environmental features, in the absence of a complete physical environment map.

Since the pure black regions (pixel value of 0) in the channel knowledge grayscale image represent areas where measurements are unavailable due to the presence of buildings, we can generate the building image by image binarization. 
Specifically, this involves preserving the black regions in the channel knowledge data while converting all other regions to white through image binarization.
This approach helps to isolate and highlight the areas occupied by buildings, providing important structural information for the model. 
Subsequently, the boundary of the building image is extracted using the Canny edge detection algorithm \cite{canny}, resulting in the building edge image. 
Therefore, the channel knowledge data, building image, and building edge image are combined to form an RGB image, which serves as the input to the model.

\section{Numerical Results}
\label{sec:5}
In this section, we first describe the datasets used for model training, including the CKMImageNet dataset~\cite{CKMImageNet} and RadioMapSeer dataset~\cite{radiomapseer}.
Furthermore, we also introduce the evaluation metrics, simulation setup, and finally present the experimental results to demonstrate the performance of the proposed approach.

\subsection{Dataset}
\subsubsection{\textbf{CKMImageNet}}
\begin{figure}[htbp]
    \centering
    \begin{subfigure}{0.49\linewidth}
        \includegraphics[width=\linewidth]{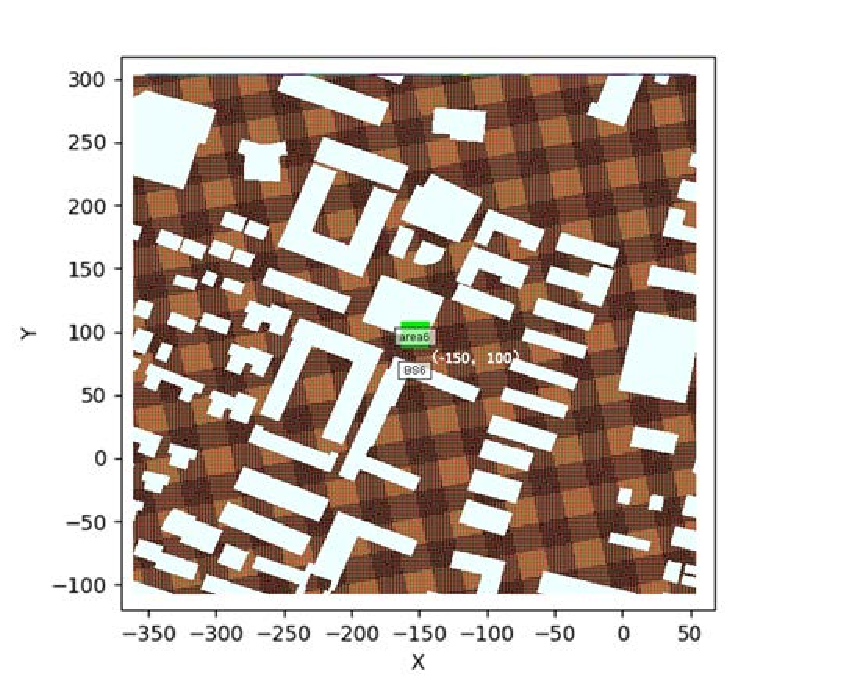}
        \caption{physical environment map}
    \end{subfigure}
    \hfill
    \begin{subfigure}{0.49\linewidth}
        \includegraphics[width=\linewidth]{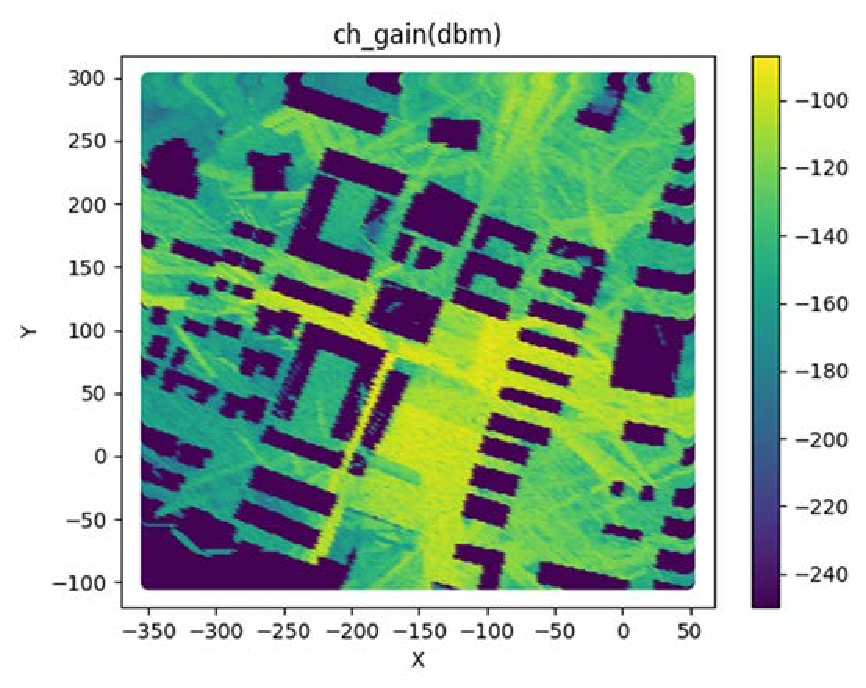}
        \caption{channel gain map}
    \end{subfigure}
    
    \vspace{0.30cm}
    
    \begin{subfigure}{0.49\linewidth}
        \includegraphics[width=\linewidth]{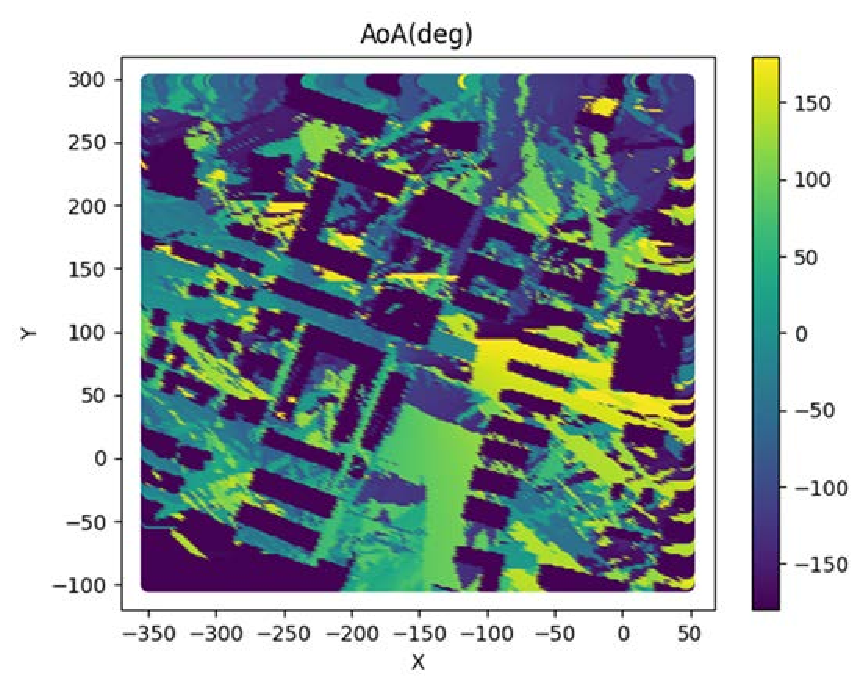}
        \caption{AoA map}
    \end{subfigure}
    \hfill
    \begin{subfigure}{0.49\linewidth}
        \includegraphics[width=\linewidth]{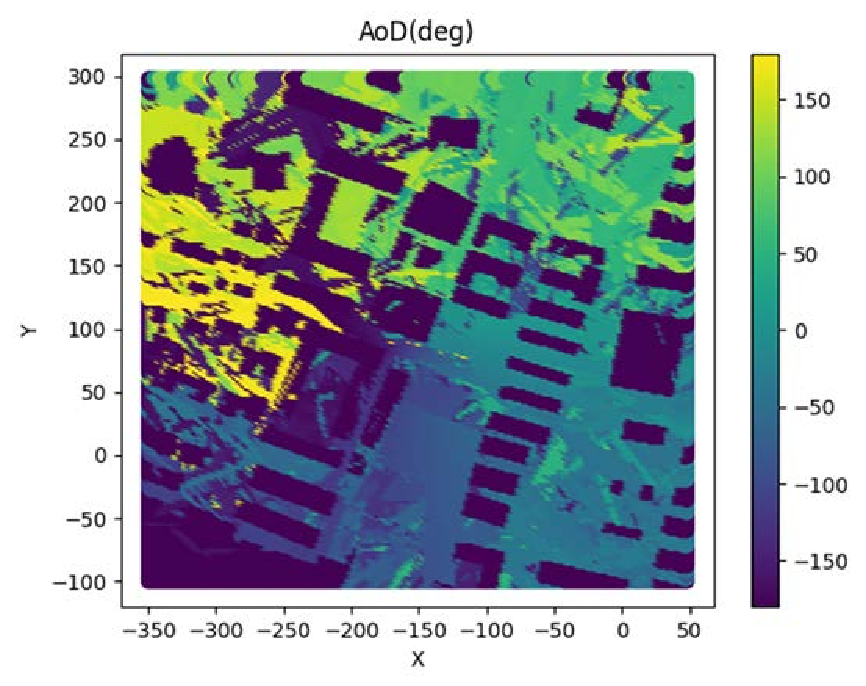}
        \caption{AoD map}
    \end{subfigure}
    \caption{An example of the physical environment map and corresponding CKMs for the CKMImageNet dataset \cite{CKMImageNet}}
    \label{graphic}
\end{figure}

The CKMImageNet dataset is a comprehensive and heterogeneous collection that combines numerical data, images of various resolutions, and environmental maps. 
It includes multiple types of CKMs, such as channel gain maps, AoA maps, AoD maps, and time of arrival (ToA) maps.
For instance, as illustrated in Fig. \ref{graphic}, a specific real-world environment map is shown, where the coordinate of BS6 is set to \(\left( {-150, 100} \right)\). 
The corresponding pseudo-colored maps of CKMs, including the channel gain map, AoA map, and AoD map, are also presented. 
These data were obtained via ray tracing by using the commerical software \textit{Wireless Insite}\textsuperscript{1}, which is renowned for its high-fidelity modeling of electromagnetic wave propagation. 
This software simulates the interactions of waves with the environment, including effects such as reflections, diffractions, and scattering.

The simulation settings involve a carrier frequency of \(28 \, \text{GHz}\), with the transmitter and receiver heights set to 10 meters and 1 meter, respectively. 
The simulation antenna is configured as an omnidirectional antenna, with the waveform set to sinusoidal waveform. The maximum number of ray reflections is $6$, and the maximum number of diffractions is $1$. Users are deployed with a uniform spacing of $2$ meters in both the $x$ and $y$ directions.
The environmental maps in the CKMImageNet dataset cover various cities, such as Beijing, Shanghai, London, and Los Angeles. 
These 3D city maps are primarily sourced from \textit{OpenStreetMap}\textsuperscript{2}, with unknown building heights being assigned a default value of 20 meters.
For simplicity, all buildings are assumed to have the same generic material property.
Currently, the CKMImageNet dataset comprises over 50,000 CKM images with resolutions of \(32 \times 32\) and \(64 \times 64\), as well as more than 30,000 CKM images with a resolution of \(128 \times 128\).
\footnotetext[1]{\url{https://www.remcom.com/wireless-insite-em-propagation-software}}
\footnotetext[2]{\url{https://www.openstreetmap.org/}}

Next, we present the relationship between the channel knowledge images and the channel knowledge data. 
For the channel gain images, the channel gain values range from \(-50\) dB to \(-250\) dB. These values are linearly mapped to the range \([0, 255]\) to generate the grayscale images required by the AI model. 
Furthermore, we assume that the values in areas corresponding to buildings are set to the minimum value of \(-250\) dB. These are mapped to a grayscale value of 0, resulting in black regions in the grayscale image.


\subsubsection{\textbf{RadioMapSeer}}
The RadioMapSeer dataset \cite{radiomapseer} consists of 700 unique radio maps, each representing a distinct geographic area with detailed information about transmitter locations and building structures. 
These maps are derived from cities such as Ankara, Berlin, Glasgow, London, and Tel Aviv, using data from OpenStreetMap. 
Each map is presented as an image with $256 \times 256$ pixels, where each pixel corresponds to a $1 \times 1$ meter square in the real-world terrain. 
The path loss maps are generated using the \textit{WinProp} software \cite{winprop}, which models the propagation of radio signals in urban environments. 
Each map in the dataset includes data for $80$ transmitter locations, with each transmitter placed at a height of $1.5$ meters. 
The transmitter power is set to $23$ dBm, and the carrier frequency is $5.9$ GHz. 
For simplicity, all buildings in the dataset are assumed to have identical material properties.

\begin{figure}[htbp]
    \centering
    \begin{subfigure}{0.35\linewidth}
        \includegraphics[width=\linewidth]{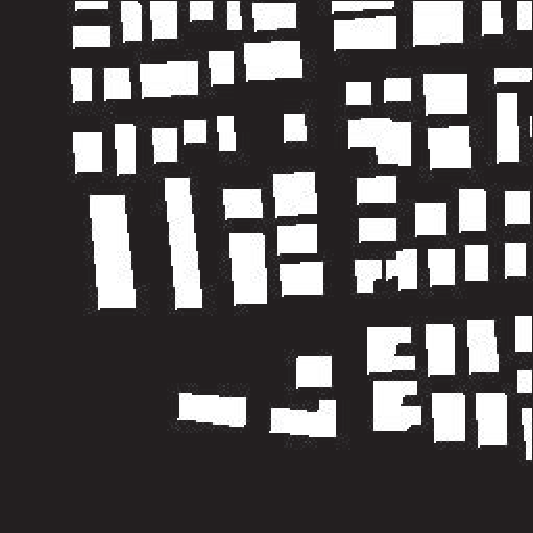}
        \caption{environment map}
    \end{subfigure}
    \hspace{0.15\linewidth} 
    \begin{subfigure}{0.35\linewidth}
        \includegraphics[width=\linewidth]{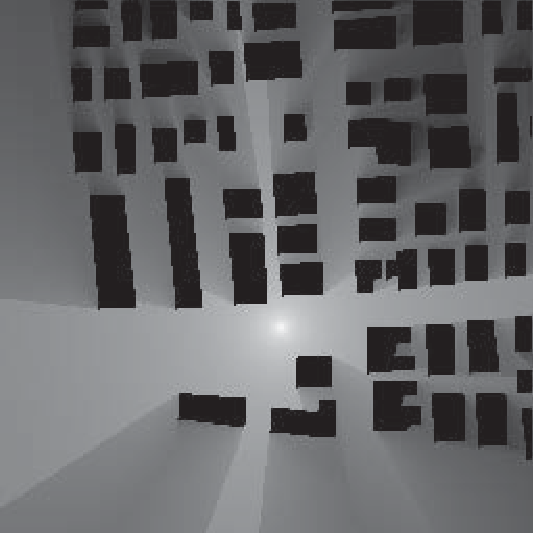}
        \caption{DPM map}
    \end{subfigure}
    
    \vspace{0.30cm}
    
    \begin{subfigure}{0.35\linewidth}
        \includegraphics[width=\linewidth]{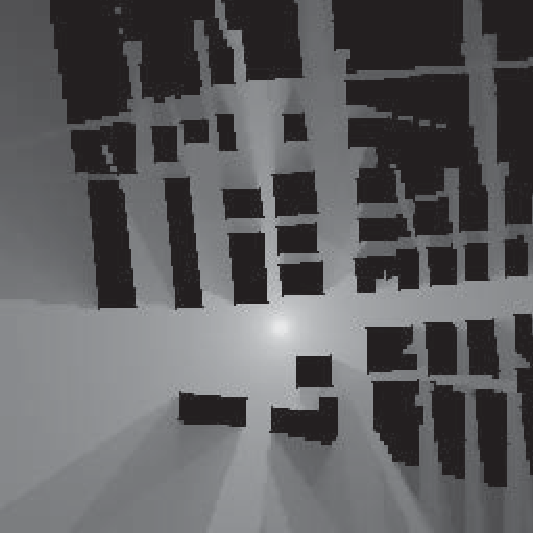}
        \caption{IRT2 map}
    \end{subfigure}
    \hspace{0.15\linewidth} 
    \begin{subfigure}{0.35\linewidth}
        \includegraphics[width=\linewidth]{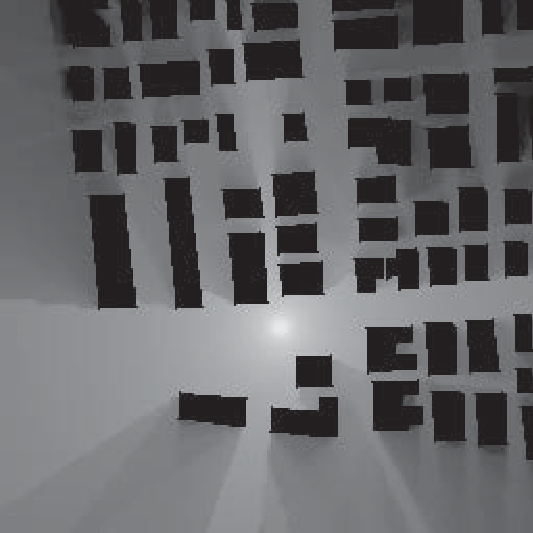}
        \caption{IRT4 map}
    \end{subfigure}
    \caption{An example of the physical environment map and corresponding path loss maps of different simulation methods for RadioMapSeer dataset \cite{radiomapseer}.}
    \label{radiomap}
\end{figure}

Two different simulation methods were employed to generate the radio maps: the dominant path model (DPM) and intelligent ray tracing (IRT). 
For this dataset,  IRT is used with two different interaction levels: IRT2 (with 2 interactions) and IRT4 (with 4 interactions). 
Fig. \ref{radiomap} displays the environment map of a specific area, along with the corresponding maps generated by DPM, IRT2, and IRT4.
In the environment map, buildings are represented in white, while the background is depicted in black.
In addition, it can be observed that IRT4 captures more intricate channel characteristics compared to the simpler representations in DPM and IRT2. 

The dataset provides simulated path loss values ranging from $-186$ dB to $-47$ dB, with a reliability threshold set at $-147$ dB, below which signal detection is considered unreliable. 
Similar to the CKMImageNet dataset, areas corresponding to buildings are assigned the minimum path loss value of $-147$ dB. 
The path loss values are then mapped to grayscale, where $0$ represents the threshold ($-147$ dB) and $255$ corresponds to the maximum path loss ($-47$ dB). 

\subsection{Performance Evaluation Metrics}
To evaluate the effectiveness of the reconstructed CKM compared to the ground-truth CKM, various performance metrics can be used. 
In addition, we focus on conducting metrics evaluation for the grayscale CKM of interest, instead of RGB images. 
These metrics include pixel-level mean squared error (MSE), normalized mean squared error (NMSE), root mean squared error (RMSE), gain-level MSE, peak signal-to-noise ratio (PSNR), structural similarity index measurement (SSIM), and Fréchet Inception Distance (FID).
\subsubsection{Pixel-level MSE}
The pixel-level MSE quantifies the overall pixel-by-pixel error between the original and reconstructed images and can be calculated as
\begin{equation}
    MS{E_{{\rm{pixel}}}} = \frac{1}{{Mlw}}\sum\limits_{m = 1}^M {\sum\limits_{i = 1}^{lw} {{{\left( {x_i^m - \hat x_i^m} \right)}^2}} } ,
    \label{mse_pixel}
\end{equation}
where ${x_i^m}$ and ${\hat x_i^m}$ denote the $i$-th normalized pixel value of the $m$-th original CKM image and reconstructed CKM image, respectively, $lw$ is the total number of pixels and $M$ represents the number of images used for evaluation.

RMSE is the square root of the pixel-level MSE, which gives the error in the same unit as the original pixel values and its definition is
\begin{equation}
	RMSE = \sqrt {MS{E_{{\rm{pixel}}}}} .
    \label{rmse_pixel}
\end{equation}

NMSE measures the MSE at the pixel level but scales the error relative to the power of the original signal. 
NMSE normalizes the error to account for the original signal's magnitude and is defined as
\begin{equation}
    NMSE = \frac{1}{M}\sum\limits_{m = 1}^M {\frac{{\sum\nolimits_{i = 1}^{lw} {{{\left( {x_i^m - \hat x_i^m} \right)}^2}} }}{{\sum\nolimits_{i = 1}^{lw} {{{\left( {x_i^m} \right)}^2}} }}} .
    \label{nmse_pixel}
\end{equation}

\subsubsection{Gain-level MSE}
Gain-level MSE computes MSE between original and reconstructed CKMs specifically for channel gain values (not pixel intensities).
It serves to directly evaluate the accuracy of channel gain reconstruction, which is essential for tasks in wireless systems where signal quality is more important than pixel-level detail.
It is defined as:
\begin{equation}
    MS{E_{{\rm{gain}}}} = \frac{1}{{Mlw}}\sum\limits_{m = 1}^M {\sum\limits_{i = 1}^{lw} {{{\left( {g_i^m - \hat g_i^m} \right)}^2}} } ,
    \label{mse_gain}
\end{equation}
where ${g_i^m}$ and ${\hat g_i^m}$ denote the $i$-th gain value of the $m$-th original CKM image and reconstructed CKM image, respectively. 
For the CKMImageNet dataset, the transformation relationship from normalized pixel values to gain values is expressed as 
\begin{equation}
	{g_i^m} = {{x_i^m} }  \times 200 - 250.
    \label{pg}
\end{equation}
Similarly, the $i$-th gain value of the $m$-th CKM image for the RadioMapSeer dataset can be obtained by
\begin{equation}
	{g_i^m} = {{x_i^m} }  \times 100 - 147.
    \label{pg_r}
\end{equation}

\subsubsection{PSNR}
PSNR is a logarithmic measure of the ratio between the maximum normalized pixel value (i.e., 1) and the error introduced by the reconstruction process, and it provides a measure of the fidelity of the reconstructed CKM images.
PSNR can be calculated as 
\begin{equation}
    PSNR = \frac{1}{M}\sum\limits_{m = 1}^M {10{{\log }_{10}}\left( {\frac{{lw}}{{\sum\nolimits_{i = 1}^{lw} {{{\left( {x_i^m - \hat x_i^m} \right)}^2}} }}} \right)} .
    \label{psnr}
\end{equation}

\subsubsection{SSIM}
SSIM is used to evaluate the perceptual quality of CKM construction by measuring structural similarities between the original and reconstructed CKMs, accounting for luminance, contrast, and texture \cite{ssim}.
It ranges from $-1$ (completely dissimilar) to $1$ (identical) and is defined as 
\begin{equation}
    SSIM = \frac{1}{M}\sum\limits_{m = 1}^M {\frac{{\left( {2{\mu _{{{\bf{x}}^m}}}{\mu _{{{{\bf{\hat x}}}^m}}} + {C_1}} \right)\left( {2{\sigma _{{{\bf{x}}^m}{{{\bf{\hat x}}}^m}}} + {C_2}} \right)}}{{\left( {\mu _{{{\bf{x}}^m}}^2 + \mu _{{{{\bf{\hat x}}}^m}}^2 + {C_1}} \right)\left( {\sigma _{{{\bf{x}}^m}}^2 + \sigma _{{{{\bf{\hat x}}}^m}}^2 + {C_2}} \right)}}} ,
    \label{ssim}
\end{equation}
where ${\mu _{{{\bf{x}}^m}}}$ and ${\mu _{{{{\bf{\hat x}}}^m}}}$ denote the mean values of the $m$-th ground-truth CKM ${\bf{x}}^m$ and the $m$-th reconstructed CKM ${{\bf{\hat x}}}^m$ respectively,  
${\sigma _{{{\bf{x}}^m}}^2}$ and ${\sigma _{{{{\bf{\hat x}}}^m}}^2}$ represent the variances of ${\bf{x}}^m$ and ${{\bf{\hat x}}}^m$ respectively, and ${{\sigma _{{{\bf{x}}^m}{{{\bf{\hat x}}}^m}}}}$ is the covariance between ${\bf{x}}^m$ and ${{\bf{\hat x}}}^m$.
${C_1}$ and ${C_2}$ are small constants used to avoid division by zero in the formula and are set as ${C_1} = {\left( {{k_1}L} \right)^2}$ and ${C_2} = {\left( {{k_2}L} \right)^2}$, where $L$ is the dynamic range of the image.

\subsubsection{FID}
FID focuses on global structure, spatial correlations, and texture patterns \cite{fid}, which are vital for complex spatial distributions of CKM.
It effectively evaluates the diversity and naturalness of reconstructed CKMs, making it ideal for tasks like inpainting or super-resolution, where structural integrity matters more than pixel-perfect accuracy.
FID measures the distance between the feature distributions of real and constructed images. 
It uses a pre-trained Inception v3 model to extract high-level features from both the real and constructed images, and FID is calculated as the Fréchet distance between the two distributions:
\begin{equation}
	FID = {\left\| {{{\bm{\mu }}_{{\rm{r}}}} - {{\bm{\mu }}_{{\rm{c}}}}} \right\|^2} + {\rm{Tr}}\left( {{{\bf{\Sigma }}_{{\rm{r}}}} + {{\bf{\Sigma }}_{{\rm{c}}}} - 2\sqrt {{{\bf{\Sigma }}_{{\rm{r}}}}{{\bf{\Sigma }}_{{\rm{c}}}}} } \right),
    \label{fid}
\end{equation}
where ${{{\bm{\mu }}_{{\rm{r}}}}}$ and ${{{\bf{\Sigma }}_{{\rm{r}}}}}$ denote the mean and covariance of features from the real images, and ${{{\bm{\mu }}_{{\rm{c}}}}}$ and ${{{\bf{\Sigma }}_{{\rm{c}}}}}$ denote the mean and covariance of features from the constructed images.


\subsection{Simulation Setup}
For CKMImageNet dataset, we use 30000 channel gain images each with size $ 128\times 128 $ as training dataset, and 2000 channel gain images each with size $ 128\times 128 $ as testing dataset. 
For DPM and IRT2 dataset, we use 40000 images each with size $ 256\times 256 $ as training dataset, and 2000 images each with size $ 256\times 256 $ as testing dataset. 
We train our conditional diffusion model tailored to a specific dataset and task through the following two steps.
Firstly, a VAE is trained to map the pixel space to the latent space.  
An AdamW optimizer is utilized with a scheduled learning rate from $5 \times {10^{ - 6}}$ to $1 \times {10^{ - 6}}$. 
Next, we train a conditional DDM using the AdamW optimizer, where prompts of the model for the super-resolution and inpainting tasks are low-resolution images and masked images, respectively.
The hyper-parameters of the trained conditional DDM for inpainting and super-resolution tasks are listed in table. \ref{Parameter_ip} and table. \ref{Parameter_sr}, respectively. 

\begin{table}[!t]
    \renewcommand{\arraystretch}{1.35}
    \caption{Training hyperparameter settings for inpainting model}
    \centering
    \setlength{\tabcolsep}{4pt} 
    \label{Parameter_ip}
    \begin{tabular}{c c | c c}
        \hline\hline                                              \\[-4mm]
        Dataset                       & CKMImageNet               & DPM / IRT2                \\
        Image size                    & $128 \times 128$          & $256 \times 256$            \\
        Batch size                    & $48$                      & $32$                        \\
        Iterations                    & $400000$                  & $300000$                    \\
        Learning rate                 & $4 \times {10^{ - 5}} \sim 4 \times {10^{ - 6}}$  &  $4 \times {10^{ - 5}} \sim 4 \times {10^{ - 6}}$  \\
        UNet feature channels              & $96$                      & $96$                        \\
        UNet channel multiplier            & $[1,2,4,8]$               & $[1,2,4,8]$                   \\
        Number of blocks              & $2$                       & $2$                          \\
		Smallest time step            & $1 \times {10^{ - 4}}$    & $1 \times {10^{ - 4}}$         \\
        \hline\hline
    \end{tabular}
\end{table}

\begin{table}[!t]
    \renewcommand{\arraystretch}{1.35}
    \caption{Training hyperparameter settings for super-resolution model}
    \centering
    \setlength{\tabcolsep}{4pt} 
    \label{Parameter_sr}
    \begin{tabular}{c c | c c}
        \hline\hline                                              \\[-4mm]
        Dataset                       & CKMImageNet               & DPM / IRT2                \\
        Image size                    & $128 \times 128$          & $256 \times 256$            \\
        Batch size                    & $16$                      & $12$                        \\
        Iterations                    & $300000$                  & $200000$                    \\
        Learning rate                 & $5 \times {10^{ - 5}} \sim 5 \times {10^{ - 6}}$  &  $5 \times {10^{ - 5}} \sim 5 \times {10^{ - 6}}$  \\
        UNet feature channels              & $128$                      & $128$                        \\
        UNet channel multiplier            & $[1,2,4,4]$               & $[1,2,4,4]$                   \\
        Number of blocks              & $2$                       & $2$                          \\
		Smallest time step            & $1 \times {10^{ - 4}}$    & $1 \times {10^{ - 4}}$         \\
        \hline\hline
    \end{tabular}
\end{table}

\subsection{Simulation Results}
In this section, we demonstrate the inpainting and super-resolution results of noisy observations through numerical simulations.

\subsubsection{inpainting task with noisy observation}
\begin{figure*}[htbp]  
    \centering
    \begin{minipage}{0.142\textwidth}  
        \centering
        \includegraphics[width=\linewidth]{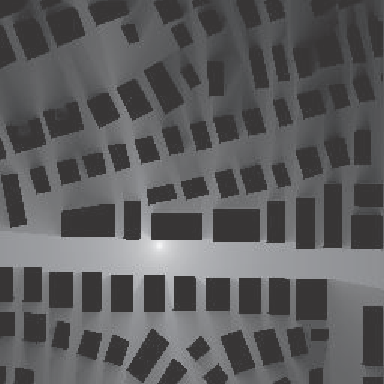}  
        \subcaption{ground truth}
    \end{minipage}
    \hfill  
    \begin{minipage}{0.142\textwidth}
        \centering
        \includegraphics[width=\linewidth]{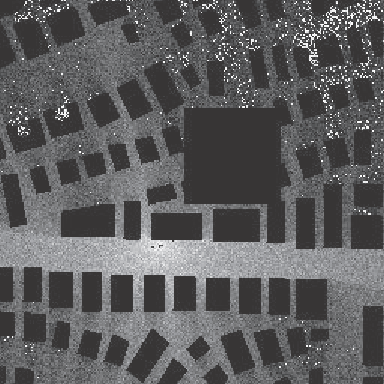}
        \subcaption{degraded image}
    \end{minipage}
    \hfill
    \begin{minipage}{0.142\textwidth}
        \centering
        \includegraphics[width=\linewidth]{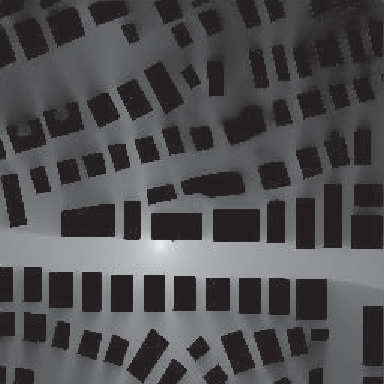}
        \subcaption{\textbf{CKMDiff}}
    \end{minipage}
    \hfill
    \begin{minipage}{0.142\textwidth}
        \centering
        \includegraphics[width=\linewidth]{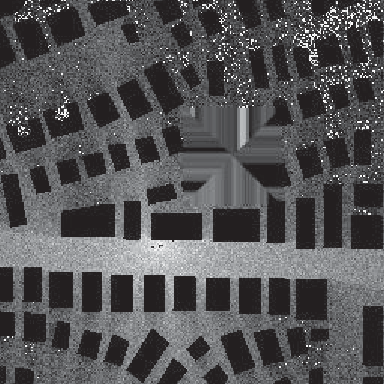}
        \subcaption{KNN}
    \end{minipage}
    \hfill
    \begin{minipage}{0.142\textwidth}
        \centering
        \includegraphics[width=\linewidth]{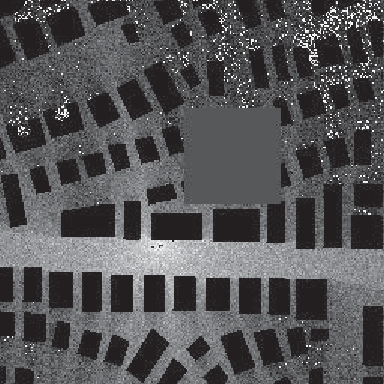}
        \subcaption{Kriging}
    \end{minipage}
    \hfill
    \begin{minipage}{0.142\textwidth}
        \centering
        \includegraphics[width=\linewidth]{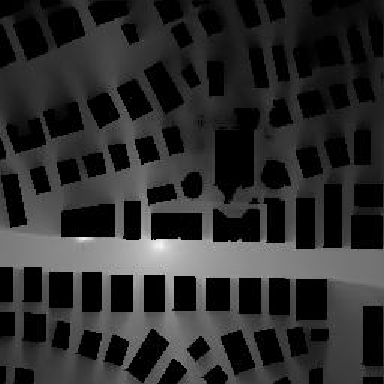}
        \subcaption{cGAN}
    \end{minipage}
    
    \caption{Visualization results of different methods in DPM for inpainting with noisy observation.}
    \label{fig:dpm_inpaint}
    \vspace*{-5pt}
\end{figure*}

\begin{figure*}[htbp]  
    \centering
    \begin{minipage}{0.142\textwidth}  
        \centering
        \includegraphics[width=\linewidth]{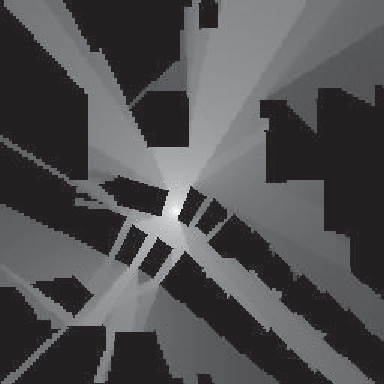}  
        \subcaption{ground truth}
    \end{minipage}
    \hfill  
    \begin{minipage}{0.142\textwidth}
        \centering
        \includegraphics[width=\linewidth]{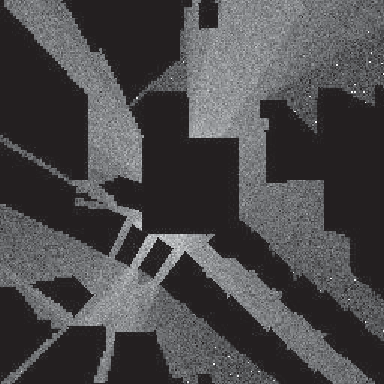}
        \subcaption{degraded image}
    \end{minipage}
    \hfill
    \begin{minipage}{0.142\textwidth}
        \centering
        \includegraphics[width=\linewidth]{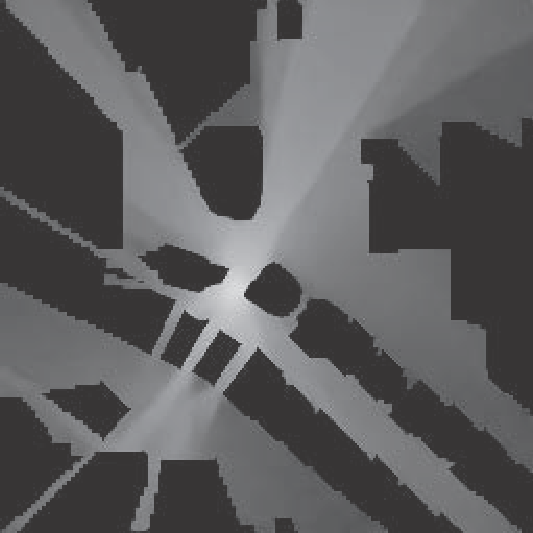}
        \subcaption{\textbf{CKMDiff}}
    \end{minipage}
    \hfill
    \begin{minipage}{0.142\textwidth}
        \centering
        \includegraphics[width=\linewidth]{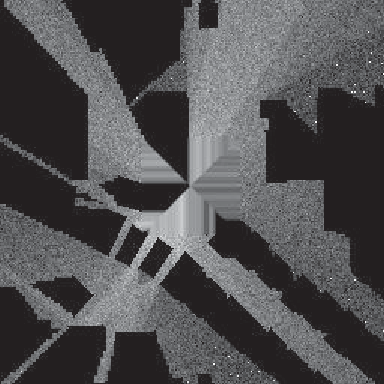}
        \subcaption{KNN}
    \end{minipage}
    \hfill
    \begin{minipage}{0.142\textwidth}
        \centering
        \includegraphics[width=\linewidth]{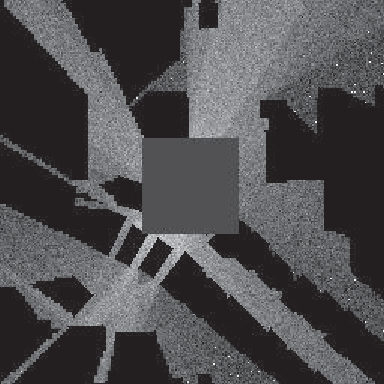}
        \subcaption{Kriging}
    \end{minipage}
    \hfill
    \begin{minipage}{0.142\textwidth}
        \centering
        \includegraphics[width=\linewidth]{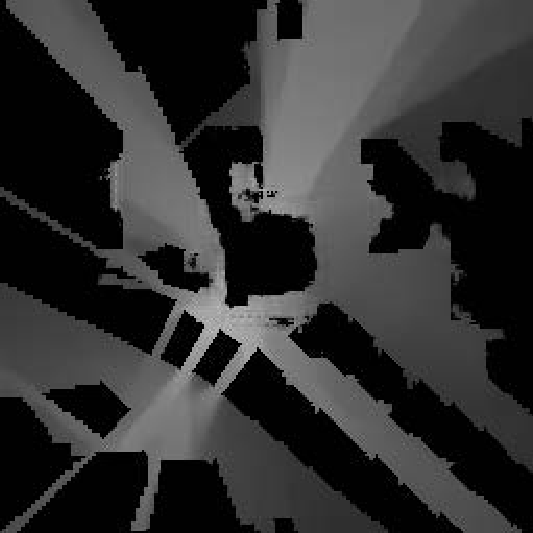}
        \subcaption{cGAN}
    \end{minipage}
    
    \caption{Visualization results of different methods in IRT2 for inpainting with noisy observation.}
    \label{fig:irt2_inpaint}
    \vspace*{-5pt}
\end{figure*}

\begin{figure*}[htbp]  
    \centering
    \begin{minipage}{0.142\textwidth}  
        \centering
        \includegraphics[width=\linewidth]{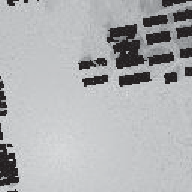}  
        \subcaption{ground truth}
    \end{minipage}
    \hfill  
    \begin{minipage}{0.142\textwidth}
        \centering
        \includegraphics[width=\linewidth]{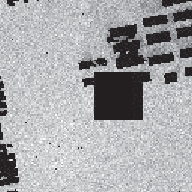}
        \subcaption{degraded image}
    \end{minipage}
    \hfill
    \begin{minipage}{0.142\textwidth}
        \centering
        \includegraphics[width=\linewidth]{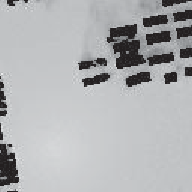}
        \subcaption{\textbf{CKMDiff}}
    \end{minipage}
    \hfill
    \begin{minipage}{0.142\textwidth}
        \centering
        \includegraphics[width=\linewidth]{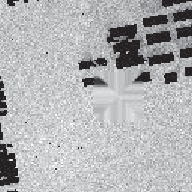}
        \subcaption{KNN}
    \end{minipage}
    \hfill
    \begin{minipage}{0.142\textwidth}
        \centering
        \includegraphics[width=\linewidth]{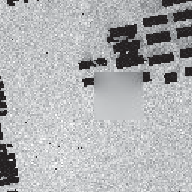}
        \subcaption{Kriging}
    \end{minipage}
    \hfill
    \begin{minipage}{0.142\textwidth}
        \centering
        \includegraphics[width=\linewidth]{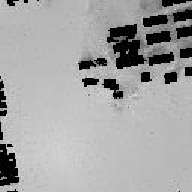}
        \subcaption{cGAN}
    \end{minipage}
    
    \caption{Visualization results of different methods in CKMImageNet for inpainting with noisy observation.}
    \label{fig:ckm_inpaint}
    \vspace*{-5pt}
\end{figure*}

Figs.~\ref{fig:dpm_inpaint}--\ref{fig:ckm_inpaint} illustrate the visualization results of inpainting with noisy observations for the datasets DPM, IRT2, and CKMImageNet, respectively, where (a) denotes the ground truth of a sample from each of the three datasets.
(b) represents the observed channel knowledge data, where for DPM and IRT2, a $64 \times 64$ region is masked on a $256 \times 256$ grayscale image, and for CKMImageNet, a $32 \times 32$ region is masked on a $128 \times 128$ grayscale image.
Gaussian noise with a mean of 0 and a standard deviation of 30 is added uniformly, with 30 referring to the grayscale intensity range (0-255).
(c) shows the reconstructed CKM obtained by our proposed method, while (e), (f), and (g) represent the results of three baseline methods: KNN, Kriging, and cGAN \cite{cgan_baseline}, respectively. 

Tables.~\ref{table:dpm_inpaint}--\ref{table:ckm_inpaint} list the evaluation metrics for inpainting with noisy observations on the DPM, IRT2, and CKMImageNet datasets, including pixel-level MSE, NMSE, RMSE, gain-level MSE, SSIM, PSNR, and FID.
The gain-level MSE is particularly important for CKM inpainting, as it measures the accuracy of reconstructed gain values, ensuring reliable communication and network optimization.
Our proposed method achieves the lowest gain-level MSE (e.g., 10.7240, 16.4865, 121.3142 for DPM, IRT2, CKMImageNet).
Besides, our proposed method also achieves the best performance in terms of PSNR and SSIM metrics.
The superior PSNR values indicate excellent noise suppression and pixel-level fidelity, while the outstanding SSIM scores reflect strong structural and geometric consistency, which is crucial for preserving multipath characteristics and spatial details in CKMs.
Additionally, the FID metric measures the statistical similarity between the original and reconstructed maps.
CKMDiff has the lowest FID value indicating closer distributional realism than baselines. 

\begin{table*}[!t]
    \renewcommand{\arraystretch}{1.5}  
    \caption{ Evaluation Metrics Comparison for DPM Inpainting with Noisy Observation}
    \centering
    \label{table:dpm_inpaint}
    \setlength{\tabcolsep}{5pt}
    \begin{tabular}{c|c|c|c|c|c|c|c}
        \hline
        \multicolumn{1}{c|}{} & \textbf{$\bm{{\rm{MS}}{{\rm{E}}_{{\rm{pixel}}}}}$ $\downarrow$} & \textbf{RMSE $\downarrow$} & \textbf{NMSE $\downarrow$} & \textbf{$\bm{{\rm{MS}}{{\rm{E}}_{{\rm{gain}}}}}$ $\downarrow$} & \textbf{PSNR $\uparrow$} &  \textbf{SSIM $\uparrow$} & \textbf{FID $\downarrow$}\\ \hline
        \textbf{KNN}  & $0.0118$ & $0.1087$ & $0.1074$ & $118.2426$ & $30.17$ & $0.45$ & $394.16 $ \\ \hline
        \textbf{Kriging}  & $0.0105$ & $0.1023$ & $0.0911$ & $104.6326$ & $29.95$ & $0.44$ & $504.09 $  \\ \hline
        \textbf{cGAN}  & $0.0041$ & $0.0643 $ & $0.0385 $ & $41.3809$ & $30.83$ & $0.92$ & $375.29 $  \\ \hline
        \textbf{CKMDiff}   & $\bf{0.0011}$ & $\bf{0.0327}$ & $\bf{0.0094}$ & $\bf{10.7240}$ & $\bf{38.70}$ & $\bf{0.96}$ & $\bf{18.08} $ \\ \hline
    \end{tabular}
\end{table*}

\begin{table*}[!t]
    \renewcommand{\arraystretch}{1.5}  
    \caption{Evaluation Metrics Comparison for IRT2 Inpainting with Noisy Observation}
    \centering
    \label{table:irt2_inpaint}
    \setlength{\tabcolsep}{5pt}
    \begin{tabular}{c|c|c|c|c|c|c|c}
        \hline
        \multicolumn{1}{c|}{} & \textbf{$\bm{{\rm{MS}}{{\rm{E}}_{{\rm{pixel}}}}}$ $\downarrow$} & \textbf{RMSE $\downarrow$} & \textbf{NMSE $\downarrow$} & \textbf{$\bm{{\rm{MS}}{{\rm{E}}_{{\rm{gain}}}}}$ $\downarrow$} & \textbf{PSNR $\uparrow$} &  \textbf{SSIM $\uparrow$} & \textbf{FID $\downarrow$}\\ \hline
        \textbf{KNN}  & $0.0060$ & $0.0777$ & $0.0762$ & $60.3773$ & $32.19$ & $0.66$ & $379.54 $ \\ \hline
        \textbf{Kriging}  & $ 0.0087$ & $0.0932$ & $0.1039$ & $ 86.9188$ & $ 30.77$  & $ 0.61$ & $  447.08$  \\ \hline
        \textbf{cGAN}  & $0.0081$ & $0.0899$ & $0.0969$ & $80.7903$ & $ 32.26$ & $0.89 $ & $549.38 $  \\ \hline
        \textbf{CKMDiff}   & $\bf{0.0016}$ & $\bf{0.0406}$ & $\bf{0.0217}$ & $\bf{16.4865}$ & $\bf{40.60}$ & $\bf{0.96}$ & $\bf{33.67} $ \\ \hline
    \end{tabular}
\end{table*}

\begin{table*}[!t]
    \renewcommand{\arraystretch}{1.5}  
    \caption{Evaluation Metrics Comparison for CKMImageNet Inpainting with Noisy Observation}
    \centering
    \label{table:ckm_inpaint}
    \setlength{\tabcolsep}{5pt}
    \begin{tabular}{c|c|c|c|c|c|c|c}
        \hline
        \multicolumn{1}{c|}{} & \textbf{$\bm{{\rm{MS}}{{\rm{E}}_{{\rm{pixel}}}}}$ $\downarrow$} & \textbf{RMSE $\downarrow$} & \textbf{NMSE $\downarrow$} & \textbf{$\bm{{\rm{MS}}{{\rm{E}}_{{\rm{gain}}}}}$ $\downarrow$} & \textbf{PSNR $\uparrow$} &  \textbf{SSIM $\uparrow$} & \textbf{FID $\downarrow$}\\ \hline
        \textbf{KNN}  & $0.0071$ & $0.0845$ & $0.0163$ & $285.4390$ & $29.45$ & $0.41$ & $ 227.23$ \\ \hline
        \textbf{Kriging}  & $0.0071$ & $0.0844$ & $0.0168$ & $284.9640$ & $29.57$  & $0.46$ & $ 247.73 $  \\ \hline
        \textbf{cGAN}  & $0.0033 $ & $0.0577 $ & $ 0.0079$ & $133.2927 $ &  $29.99$ & $ 0.79$ & $ 317.80$  \\ \hline
        \textbf{CKMDiff}   & $\bf{0.0030}$ & $\bf{0.0551}$ & $\bf{0.0068}$ & $\bf{121.3142}$ & $ \bf{34.07}$ & $ \bf{0.81}$ & $\bf{64.19} $ \\ \hline
    \end{tabular}
\end{table*}

\begin{table*}[!t]
    \renewcommand{\arraystretch}{1.5}  
    \caption{Evaluation Metrics Comparison for DPM Super-resolution with Noisy Observation}
    \centering
    \label{table:dpm_sr}
    \setlength{\tabcolsep}{5pt}
    \begin{tabular}{c|c|c|c|c|c|c|c}
        \hline
        \multicolumn{1}{c|}{} & \textbf{$\bm{{\rm{MS}}{{\rm{E}}_{{\rm{pixel}}}}}$ $\downarrow$} & \textbf{RMSE $\downarrow$} & \textbf{NMSE $\downarrow$} & \textbf{$\bm{{\rm{MS}}{{\rm{E}}_{{\rm{gain}}}}}$ $\downarrow$} & \textbf{PSNR $\uparrow$} &  \textbf{SSIM $\uparrow$} & \textbf{FID $\downarrow$}\\ \hline
        \textbf{Bilinear}  & $0.0019$ & $0.0439$ & $0.0125$ & $19.2838$ & $34.60$ & $0.84$ & $375.14 $ \\ \hline
        \textbf{Bicubic}  & $0.0017$ & $0.0416$ & $0.0112$ & $17.3038$ & $34.15$  & $0.83$ & $389.96 $  \\ \hline
        \textbf{cGAN}  & $0.0009 $ & $ 0.0307$ & $ 0.0061$ & $9.3992 $ & $40.39$ & $0.95$ & $58.26$  \\ \hline
        \textbf{CKMDiff}   & $\bf{0.0002}$ & $\bf{0.0146}$ & $\bf{0.0014}$ & $\bf{2.1304}$ & $\bf{42.09}$ & $\bf{0.98}$ & $\bf{14.94} $ \\ \hline
    \end{tabular}
\end{table*}

\begin{table*}[!t]
    \renewcommand{\arraystretch}{1.5}  
    \caption{Evaluation Metrics Comparison for IRT2 Super-resolution with Noisy Observation}
    \centering
    \label{table:irt2_sr}
    \setlength{\tabcolsep}{5pt}
    \begin{tabular}{c|c|c|c|c|c|c|c}
        \hline
        \multicolumn{1}{c|}{} & \textbf{$\bm{{\rm{MS}}{{\rm{E}}_{{\rm{pixel}}}}}$ $\downarrow$} & \textbf{RMSE $\downarrow$} & \textbf{NMSE $\downarrow$} & \textbf{$\bm{{\rm{MS}}{{\rm{E}}_{{\rm{gain}}}}}$ $\downarrow$} & \textbf{PSNR $\uparrow$} &  \textbf{SSIM $\uparrow$} & \textbf{FID $\downarrow$}\\ \hline
        \textbf{Bilinear}  & $0.0035$ & $0.0588$ & $0.0654$ & $ 34.6309$ & $32.97$ & $0.77$ & $288.75 $ \\ \hline
        \textbf{Bicubic}  & $0.0029$ & $0.0542$ & $0.0555$ & $29.3845$ & $33.32$  & $0.80$ & $272.40 $  \\ \hline
        \textbf{cGAN}  & $0.0027 $ & $0.0515 $ & $ 0.0457$ & $26.5536 $ & $37.78$ & $0.89$ & $63.82 $  \\ \hline
        \textbf{CKMDiff}  & $\bf{0.0007}$ & $\bf{0.0255}$ & $\bf{0.0123}$ & $\bf{6.5136}$ & $\bf{39.24}$ & $\bf{0.96}$ & $ \bf{20.45} $ \\ \hline
    \end{tabular}
\end{table*}

\begin{table*}[!t]
    \renewcommand{\arraystretch}{1.5}  
    \caption{Evaluation Metrics Comparison for CKMImageNet Super-resolution with Noisy Observation}
    \centering
    \label{table:ckm_sr}
    \setlength{\tabcolsep}{5pt}
    \begin{tabular}{c|c|c|c|c|c|c|c}
        \hline
        \multicolumn{1}{c|}{} & \textbf{$\bm{{\rm{MS}}{{\rm{E}}_{{\rm{pixel}}}}}$ $\downarrow$} & \textbf{RMSE $\downarrow$} & \textbf{NMSE $\downarrow$} & \textbf{$\bm{{\rm{MS}}{{\rm{E}}_{{\rm{gain}}}}}$ $\downarrow$} & \textbf{PSNR $\uparrow$} &  \textbf{SSIM $\uparrow$} & \textbf{FID $\downarrow$}\\ \hline
        \textbf{Bilinear}  & $0.0109$ & $0.1046$ & $0.0327$ & $437.5732$ & $30.90 $ & $0.63$  & $486.94 $ \\ \hline
        \textbf{Bicubic}  & $0.0090$ & $0.0949$ & $0.0269$ & $360.4962$ & $31.05$  & $0.66$  & $  433.92$  \\ \hline
        \textbf{cGAN}  & $0.0121$ & $ 0.1101$ & $0.0406 $ & $485.0265 $ & $32.53$ & $0.68$ & $ 144.90$  \\ \hline
        \textbf{CKMDiff}   & $\bf{0.0041}$ & $\bf{0.0637}$ & $\bf{0.0121}$ & $\bf{162.5268}$ & $\bf{32.90}$ & $\bf{0.78}$  & $ \bf{114.17} $ \\ \hline
    \end{tabular}
\end{table*}

\subsubsection{super-resolution task with noisy observation}
\begin{figure*}[htbp]  
    \centering
    \begin{minipage}{0.142\textwidth}  
        \centering
        \includegraphics[width=\linewidth]{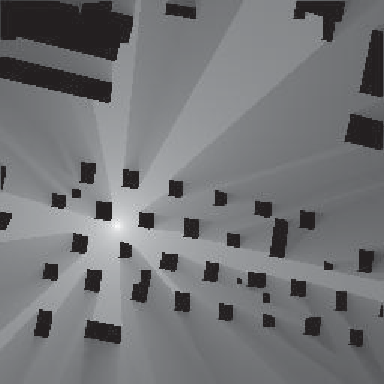}  
        \subcaption{ground truth}
    \end{minipage}
    \hfill  
    \begin{minipage}{0.145\textwidth}
        \centering
        \includegraphics[width=\linewidth]{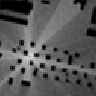}
        \subcaption{degraded image}
    \end{minipage}
    \hfill
    \begin{minipage}{0.142\textwidth}
        \centering
        \includegraphics[width=\linewidth]{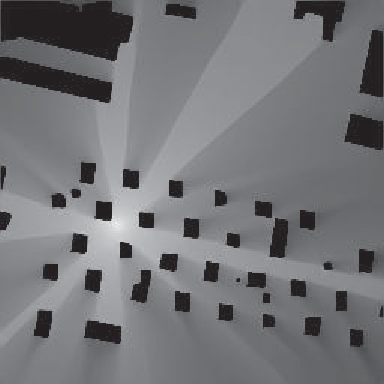}
        \subcaption{\textbf{CKMDiff}}
    \end{minipage}
    \hfill
    \begin{minipage}{0.142\textwidth}
        \centering
        \includegraphics[width=\linewidth]{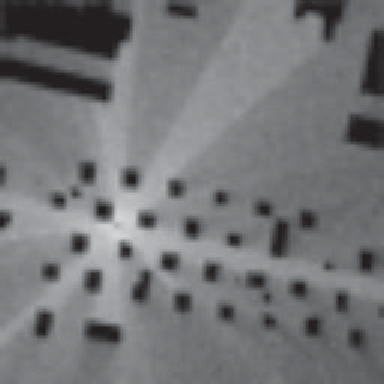}
        \subcaption{Bilinear}
    \end{minipage}
    \hfill
    \begin{minipage}{0.142\textwidth}
        \centering
        \includegraphics[width=\linewidth]{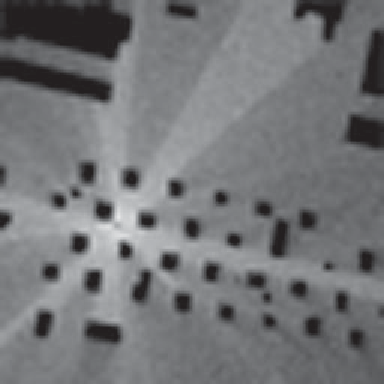}
        \subcaption{Bicubic}
    \end{minipage}
    \hfill
    \begin{minipage}{0.142\textwidth}
        \centering
        \includegraphics[width=\linewidth]{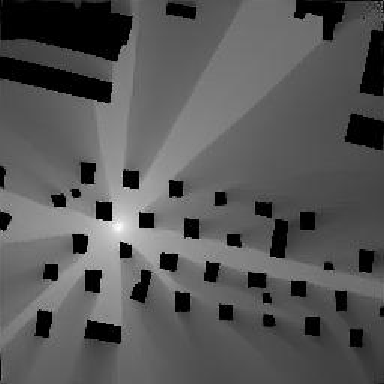}
        \subcaption{cGAN}
    \end{minipage}
    
    \caption{Visualization results of different methods in DPM for super-resolution with noisy observation.}
    \label{fig:dpm_sr}
    \vspace*{-5pt}
\end{figure*}

\begin{figure*}[htbp]  
    \centering
    \begin{minipage}{0.142\textwidth}  
        \centering
        \includegraphics[width=\linewidth]{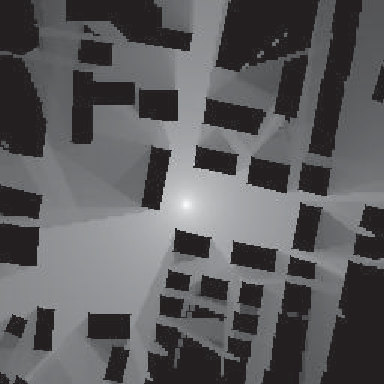}  
        \subcaption{ground truth}
    \end{minipage}
    \hfill  
    \begin{minipage}{0.145\textwidth}
        \centering
        \includegraphics[width=\linewidth]{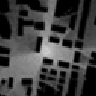}
        \subcaption{degraded image}
    \end{minipage}
    \hfill
    \begin{minipage}{0.142\textwidth}
        \centering
        \includegraphics[width=\linewidth]{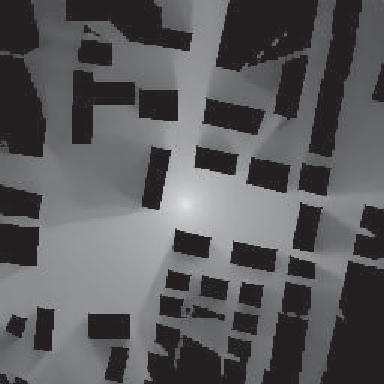}
        \subcaption{\textbf{CKMDiff}}
    \end{minipage}
    \hfill
    \begin{minipage}{0.142\textwidth}
        \centering
        \includegraphics[width=\linewidth]{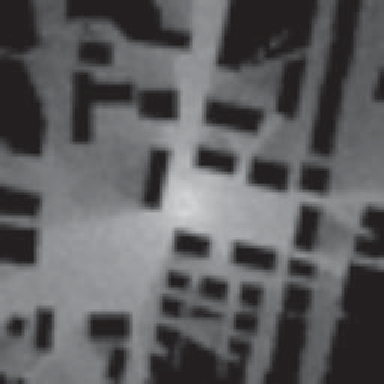}
        \subcaption{Bilinear}
    \end{minipage}
    \hfill
    \begin{minipage}{0.142\textwidth}
        \centering
        \includegraphics[width=\linewidth]{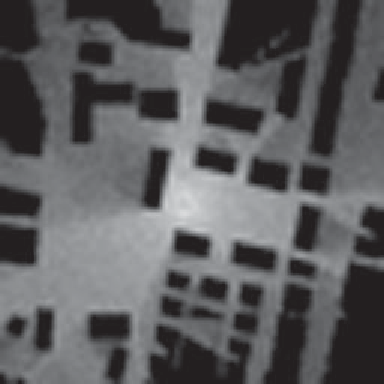}
        \subcaption{Bicubic}
    \end{minipage}
    \hfill
    \begin{minipage}{0.142\textwidth}
        \centering
        \includegraphics[width=\linewidth]{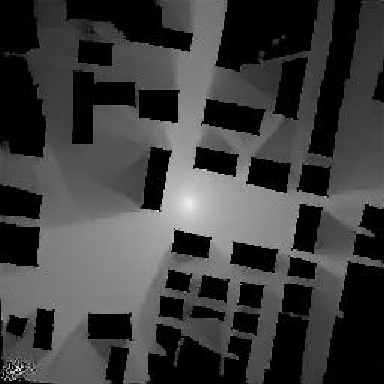}
        \subcaption{cGAN}
    \end{minipage}
    
    \caption{Visualization results of different methods in IRT2 for super-resolution with noisy observation.}
    \label{fig:irt2_sr}
    \vspace*{-5pt}
\end{figure*}

\begin{figure*}[htbp]  
    \centering
    \begin{minipage}{0.142\textwidth}  
        \centering
        \includegraphics[width=\linewidth]{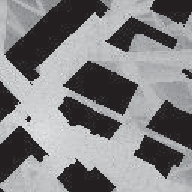}  
        \subcaption{ground truth}
    \end{minipage}
    \hfill  
    \begin{minipage}{0.1485\textwidth}
        \centering
        \includegraphics[width=\linewidth]{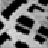}
        \subcaption{degraded image}
    \end{minipage}
    \hfill
    \begin{minipage}{0.142\textwidth}
        \centering
        \includegraphics[width=\linewidth]{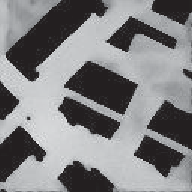}
        \subcaption{\textbf{CKMDiff}}
    \end{minipage}
    \hfill
    \begin{minipage}{0.142\textwidth}
        \centering
        \includegraphics[width=\linewidth]{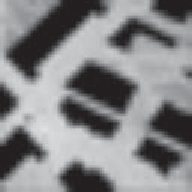}
        \subcaption{Bilinear}
    \end{minipage}
    \hfill
    \begin{minipage}{0.142\textwidth}
        \centering
        \includegraphics[width=\linewidth]{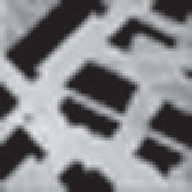}
        \subcaption{Bicubic}
    \end{minipage}
    \hfill
    \begin{minipage}{0.142\textwidth}
        \centering
        \includegraphics[width=\linewidth]{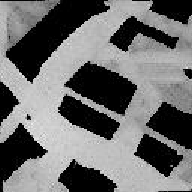}
        \subcaption{cGAN}
    \end{minipage}
    
    \caption{Visualization results of different methods in CKMImageNet for super-resolution with noisy observation.}
    \label{fig:ckm_sr}
    \vspace*{-5pt}
\end{figure*}

Figs.~\ref{fig:dpm_sr}--\ref{fig:ckm_sr} illustrate the visualization examples of super-resolving noisy observations for DPM, IRT2, and CKMImageNet, respectively.
In these figures, (a) represents the ground truth, (b) depicts the low-resolution CKM downsampled by a factor of 4 with added Gaussian noise, following ${\cal N}\left( {0,{{30}^2}} \right)$, and (c) shows the high-resolution CKM reconstructed using our proposed CKMDiff.
For comparison, baseline methods including bilinear interpolation, bicubic interpolation, and cGAN are displayed in (d), (e), and (f), respectively.
Tables.~\ref{table:dpm_sr}--\ref{table:ckm_sr} provide the quantitative performance evaluations for the three datasets.
It is evident that CKMDiff significantly outperforms the baseline approaches.
Specifically, the gain-level MSE achieved by our method is 2.1304, 6.5136, and 162.5268 for DPM, IRT2, and CKMImageNet, respectively.
These results show that the proposed diffusion model can effectively learn the intrinsic relationships within channel knowledge, enabling more accurate reconstruction and denoising.

Furthermore, it is observed that the dataset DPM leads to smaller errors than IRT2 and CKMImageNet, both in terms of CKM inpainting and super-resolution.
This is because DPM and IRT2 consider simplified scenarios: DPM assumes identical building height and it only models the dominant path, while IRT2 incorporates ray tracing with 2-ray interactions.
In contrast, CKMImageNet is constructed using real-world 3D physical environments and employs ray tracing with more complex interactions, specifically 6 reflections and 1 diffraction, which better reflects the realistic electromagnetic propagation conditions.
While this complexity makes CKMImageNet more challenging to learn, it also provides a more accurate representation of real-world scenarios, highlighting the robustness and generalization capability of our proposed diffusion model in handling diverse and complex CKMs.

\section{Conclusion}
\label{sec:6}
In this paper, we proposed CKMDiff, a conditional diffusion model framework for recovering high-fidelity CKM from corrupted observable data.
Firstly, we designed a data feature extraction mechanism that incorporates partial building geometry features from observable data, enabling the model to better learn the intrinsic correlations between CKM patterns and physical environments.
Furthermore, we developed task-specific diffusion models tailored for inpainting and super-resolution scenarios, where conditional prompts are constructed from noisy and incomplete channel knowledge data and noisy low-resolution channel knowledge data, respectively.
Simulations demonstrate the superiority of CKMDiff over baseline methods across the CKMImageNet and RadioMapSeer datasets, achieving state-of-the-art performance. 


\bibliographystyle{IEEEtran}
\bibliography{reff}

\end{document}